# The Joint Dark Energy Mission (JDEM) / Omega

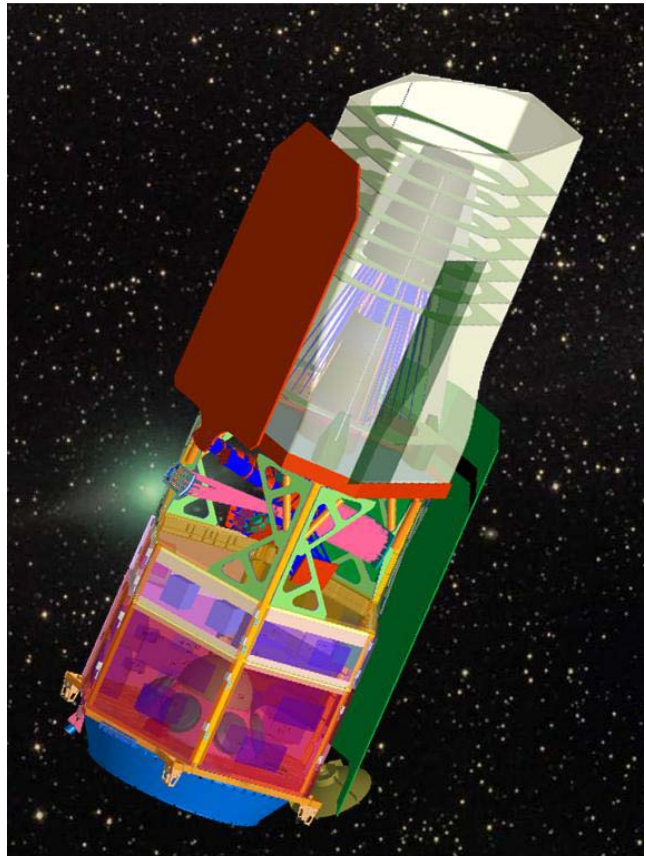

Submitted By:

Neil Gehrels
JDEM/Omega Project Scientist
Goddard Space Flight Center (GSFC)
301-286-6546
neil.gehrels@nasa.gov

# EXECUTIVE SUMMARY

The primary scientific objective of the Joint Dark Energy Mission (JDEM) is to determine the nature of dark energy in the Universe by measuring the expansion history and the growth rate of large scale structure. These observables probe the dark energy equation of state and test the validity of General Relativity. The cause of the accelerated expansion of the Universe is one of the most important and profound scientific questions of our time, and JDEM has the best capabilities of any proposed experiment to answer them. JDEM is designed to perform the critical observations that are difficult or impossible from the ground.

The JDEM/Omega concept described herein is a powerful wide-field NIR survey mission, orders of magnitude more sensitive than anything previous. It will enable a major step forward in dark energy understanding in addition to providing an ancillary data set of great value to the astronomical community. The JDEM scientific objectives are:

**Cosmic Acceleration Objective:** Determine the cosmic equation of state and its change with time to a factor of at least 10 better than current (Stage II) experiments as defined by the Dark Energy Task Force "Figure of Merit" (FoM) (Albrecht et al. arXiv 0901.0721).

**Cosmic Growth of Structure Objective:** Determine the cosmic growth of structure to a factor of at least 100 better than current (Stage II) experiments as measured by the Figure of Merit Science Working Group gamma parameter. (Goal for JDEM/Omega)

**Sky Survey Objective:** Perform a spectroscopic and multi-band high-resolution imaging survey in the NIR to obtain redshifts for $>10^8$ galaxies and images for $>10^9$ galaxies; a factor of $>100$ more than currently available.

JDEM is a mission concept collaboratively developed by NASA and the Department of Energy (DOE), with substantial community-based input. A signed MOU is in place to define the NASA-DOE collaboration. As requested in the RFI, this response describes a smaller version of the mission (~$1.2B FY09 total cost), called JDEM/Omega, while a companion response describes a larger version called JDEM/DECS (~$1.5B FY09 total cost). JDEM/DECS has CCD and HgCdTe instruments, while JDEM/Omega has a HgCdTe instrument covering both NIR and visible bands. Both provide data to enable an order of magnitude improvement in measuring the equation of state parameters of the Universe. Three different, powerful and complementary observational techniques are employed: Baryon Acoustic Oscillations (BAO), Type Ia Supernovae (SNe) and Weak Lensing (WL). JDEM/DECS performs

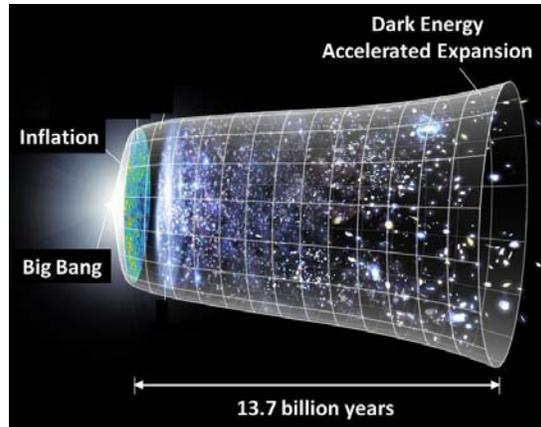

*Figure 1 - An illustration of how the Universe's expansion rate is increasing in the current epoch. Astrophysicists have no understanding of what causes this acceleration. JDEM is designed to make the necessary measurements for a breakthrough in our knowledge of the "dark energy" that drives the acceleration. (Figure by NASA/WMAP Science Team)*

the WL galaxy shape measurements in the visible with CCDs while JDEM/Omega performs it in the NIR with HgCdTe's. The smaller JDEM/Omega version addresses the Astro2010 EOS questions about particular techniques driving mission cost and complexity. The WL shape measurement is a goal for JDEM-Omega.

Over the five year mission, JDEM/Omega will execute a wide-field NIR, 3-D spectroscopic survey of emission-line galaxies and a wide-field imaging and photo-z NIR survey of field galaxies. These surveys will be two orders of magnitude larger than any currently available and will provide enormous catalogs of astrophysical objects for many communities, ranging from solar system to galaxies/clusters to cosmology. JDEM will be synergistic with JWST, overlapping in time and providing an exhaustive set of sources for JWST deep follow-up. To accomplish these surveys, the JDEM/Omega reference mission uses a 1.5 m telescope to feed a single instrument. The instrument contains three channels, an imager and two identical, slitless spectrometers. All three channels use HgCdTe devices. The imager covers 0.4-2.0 $\mu$m with a pixel scale of 0.18 arcsec/pixel and the spectrometers cover 1.1-2.0 $\mu$m with a pixel scale of 0.37 arcsec/pixel. An L2 orbit minimizes concerns with stray light from the Earth or moon, provides an unobstructed view of the sky and a thermally stable environment. No new technologies are required to build JDEM, which can be ready for launch in 2017.

# 1. SCIENCE

1. Describe the measurements required to fulfill the scientific objectives expected to be achieved by your activity.

JDEM/Omega performs a comprehensive survey of the moderate redshift universe in the near infrared (NIR). The combination of multi-color imaging data and spectroscopic redshift measurements will enable JDEM/Omega to probe dark energy via three independent techniques: baryon acoustic oscillations (BAO), supernovae (SN), and weak lensing (WL). *Figure 2* summarizes the flowdown from each method's measurement requirements to data set and design (instrument and operations) requirements.

**Baryon Acoustic Oscillations:** Coherent oscillations in the baryon-photon fluid prior to recombination (380,000 years after the Big Bang) imparted a series of peaks in the power spectrum of both the CMB radiation and the large-scale galaxy distribution on the scale of the sound horizon (the distance a sound wave travels by the epoch of recombination). Since the physical scale of the sound horizon is very well known, it forms a standard ruler which can be used to measure the angular diameter distance as a function of redshift, and hence to probe the expansion history of the universe.

The CMB signature has been observed with high precision and accuracy by WMAP, while the galaxy signature has been detected at low redshift ($z$~0.35) by the SDSS and 2dF sky surveys. The JDEM/Omega BAO measurement requirements are to determine the angular diameter distance over the important redshift range $0.7 < z < 2.0$ (the range over which the expansion of the Universe transitioned from decelerating to accelerating) to within a factor of 2 of the cosmic variance limit. This requires a spectroscopic redshift survey of ~1x10⁸ emission line galaxies over ~20,000 deg² of sky, with a redshift accuracy of $\sigma_z \lesssim 0.001(1+z)$, and a line mis-identification rate of ≤10%. Required observations comprise single color imaging data (S/N>10 for H_AB<23.5) for source identification and slitless spectroscopy data ($1.1 < \lambda < 2.0$ μm, $R_\Theta = 200$-240 arcsec, S/N>6.5 for a 1.6x10⁻¹⁶ erg/cm²-s line at 2.0 um) taken over 4 roll angles (i.e., dispersion directions) to mitigate source confusion and maintain redshift accuracy.

Since the BAO signal is based on position and wavelength information, photometric calibration and dust corrections are not stringent requirements. Nonlinear astrophysical effects smooth the peaks to first order and shift their positions to second order. The former only affects sensitivity and is accounted for in FoM forecasts while the latter affects the distance scale at the sub-percent level - a few sigma at JDEM sensitivity - and will need to be corrected (Percival et al., SDSS BAO paper, arXiv:0907.1660).

**Supernovae:** Type Ia supernovae are standardizable candles that allow one to measure the luminosity distance as a function of redshift, and hence the expansion history. The key to realizing the cosmological potential of supernovae is to obtain accurately calibrated light curves with multi-color information to measure extinction. Reliable typing and redshift information are also necessary.

The JDEM/Omega SNe measurement requirements are to provide ≥~8 deg²-yr of field monitoring to obtain the brightness and redshift of ≥1500 Type Ia SNe, in the redshift range $z$ = 0.2-1.3, with a sample size of ~150 SNe per $\Delta z$ = 0.1 bin, a redshift precision of $\sigma_z < 0.005(1+z)$, and a distance error of $\sigma_L \leq 0.007$ per $\Delta z$ = 0.1 bin. Required observations comprise multi-band imaging (0.4-1.7μm) and slitless prism spectroscopy (R=75, 2-pixel) of fields near the ecliptic poles that can be observed continuously throughout one or more years. Each field will be observed at intervals of ≤5 days. The broad-band photometry, with an absolute photometric accuracy of ≤1%, will be used to construct multi-color light curves of the SNe, and the spectra obtained near the peak of each light curve will provide the type and redshift of each SN. The light curves are transformed to the rest frame and fitted to determine the extinction and "stretch"-corrected peak magnitude, from which the distance may be inferred.

**Weak Lensing:** The WL signal probes the matter distribution along the line of site, which is sensitive to both the expansion history and the growth rate of structure. As photons from distant galaxies stream toward us, they are deflected by the gravitational fields arising from the intervening matter in the Universe. In the process, slight distortions (shear) are impressed upon the images of the source galaxies. By resolving each galaxy's image adequately to measuring its shape (ellipticity), we can reconstruct the underlying matter distribution and its growth along the line of sight. The intrinsic, unlensed shapes of the galaxies are unknown, producing an unavoidable source of measurement uncertainty. The scientifically desired lensing statistics can, however, be determined to high accuracy because of the billions of galaxies available to be measured across the sky.

The JDEM/Omega WL measurement requirements are to provide ~10,000 deg² of sky coverage over which ≥30 galaxies/arcmin² are resolved to deliver ~1x10⁹ ga-laxy images with additive shear errors of $\leq 3 \times 10^{-4}$, multiplicative shear errors of $\leq 10^{-3}$, and a photo-$z$ error distribution of $\leq 0.04(1+z)$.

Required observations comprise multi-band NIR (0.85-1.7 $\mu$m) near-Nyquist imaging of $\leq 25.5$ mag ($z<\sim 3$) galaxies to a S/N $\geq 25$, along with $\sim R=75$ (2-pixel) spectroscopy of $10^5$ galaxies to provide a photo-$z$ training set with a redshift accuracy of $\sigma_z \leq 0.01(1+z)$. In addition, 4 bands of visible imaging ($\sim 0.4$-$\sim 0.85$ $\mu$m) over the sky observed by JDEM/Omega for WL will need to be provided by ground observations similar in scope to the 4m Blanco Dark Energy Survey to support photo-$z$ determinations.

**NIR Sky Survey:** Over the five year mission, JDEM/Omega will execute a wide-field, NIR spectroscopic survey of $>10^8$ emission-line galaxies in the redshift range 0.7<z<2.0, and a multi-band, wide-field imaging and photo-$z$ survey of $>10^9$ field galaxies. These surveys will be two orders of magnitude larger than any currently available and will provide enormous catalogs of astrophysical objects for solar system studies, galaxies/cluster studies, and cosmology.

The cosmological applications alone (beyond dark energy) are many: the matter power spectrum measurement will complement the BAO signal in constraining cosmological parameters; redshift-space distortions measured by JDEM can probe the growth of structure; higher-order moments in the galaxy distribution will provide a new probe a primordial nongaussianity, an important test of inflation; the large-scale structure template can be correlated with the CMB anisotropy to probe for missing baryons (via the kinetic SZ effect) and with the CMB lensing signal to probe galaxy bias. The NIR imaging data, in concert with optical imaging, will vastly improve photo-$z$ accuracy (critical for weak lensing) and provide a unique database of very red objects, e.g. dwarf stars and high-$z$ galaxies. JDEM will be synergistic with JWST, overlapping in time and providing an exhaustive source for JWST deep follow-up.

2. *Describe the technical implementation you have selected, and how it performs the required measurements.*

*Figure 2* provides a general flowdown from the measurement requirements for each method to derivative data set, instrumentation, and operations requirements. The following summarizes the overall technical implementation and then briefly addresses selected technical implementation aspects of each method.

**Technical Implementation Overview:** JDEM/Omega utilizes a modestly cool (~243K, to ena-

ble background limited observations at 2 $\mu$m) 1.5m diameter aperture focal telescope to simultaneously feed 3 separate focal plane assemblies. Telescope collimators feed 2 separate afocal FOVs to 2 oppositely-dispersed reimaging R=270-327 (2-pixel) spectrometer cameras (1.1-2 $\mu$m; ~0.26 deg$^2$ each; 0.37 arcsec pixels; ~180 K pupil masks), while a focal FOV is fed directly to a single imager (0.4-2 $\mu$m; ~0.25 deg$^2$; 0.18 arcsec pixels; ~180 K pupil mask) with a 7-position filter wheel that includes an R=75 (2-pixel) disperser and a "dark" position. Standard 2.5 $\mu$m JWST HgCdTe detector material is used in all of the focal planes, with 2$\mu$m bandpass cutoff filters being used in the spectrometers and on the low-resolution disperser. The HgCdTe material has an acceptable QE (>0.6) down to 0.4 $\mu$m, allowing JDEM/Omega to make SNe-required visible observations without the use of CCDs.

Based on WL shape measurement accuracy concerns, the focal telescope form was chosen to minimize the number of optical elements in the imaging optical path (4, all reflective, in the telescope, and none in the instrument imaging channel except for the filters/disperser on the filter wheel). Refractive optics were chosen for the spectrometer channels primarily based on packaging volume and cold focal plane positioning considerations.

JDEM/Omega is placed in a libration point orbit about the Sun-Earth L2 point to provide a thermally stable observing platform with excellent passive cooling accommodation that can achieve a high observing efficiency due to minimal observational constraints (stray light, occultations, eclipses, etc.). The field of regard is 80° to 120° off the Sun, and inertially fixed pointing directions near the ecliptic poles can be maintained for up to ~90 days. Details related to the design and operations concepts can be found in the Technical Implementation and Mission Design sections.

**Measurement Flowdown Overview:** With an overall mission lifetime constraint of 5 years, sky coverage and cadence requirements for each method had to be satisfied by a combination of three key design parameters: integration time, FOV size, and observing efficiency. The integration time in combination with S/N, bandpass, detector choice, and system PSF requirements drove effective area and noise requirements; the FOV size in combination with system PSF, PSF sampling, and detector requirements drove the optical and focal plane layouts, and the observing efficiency drove slew/settle times, the gimbaled antenna, and sky mapping strategies. Key technical implementations unique to each method including ground vs. space considera-tions are briefly summarized below.

**BAO Implementation Specifics:** The BAO survey requires single-color imaging of modest depth for source identification and Hα (0.6563 μm) emission-line spectroscopy (1.1-2.0 μm) for redshift determination. In order to meet sky coverage requirements, the BAO demands slitless spectroscopy, including at least 4 different dispersion directions (or rolls; two nearly opposed), to mitigate source confusion and redshift errors due to offsets between galaxy line and continuum emissions. The required spectrometer FOV could not be obtained in one focal plane assembly while meeting PSF requirements, so two separate spectrometer channels are provided. This made it convenient to meet the opposed dispersion requirement by dispersing the two spectrometer channels in opposite directions, eliminating any need to revisit an observed field at least 4-6 months after an initial observation.

A fast "BAO-only" survey observes the sky twice at slightly different roll angles. The speed of this mode (6900 deg²/yr) results from the total required integration time of 1800 s being accumulated in parallel on both of the spectrometer FOVs. Note that when WL data is being acquired, BAO spectroscopy with the requisite roll angles and total spectroscopic integration times of up to 3600 s is also acquired (this is the "WL/BAO-combined" survey mode). This much deeper spectroscopy provides a better characterization of false (non-Hα) line interloper rates, helping to tune line identification strategies for the full survey.

The JDEM/Omega BAO survey would extend BAO measurements well beyond the ground-based BOSS experiment at $z < 0.7$, WiggleZ at $z\sim0.9$, and the 400-fiber FMOS spectrograph on Subaru at $0.5 \le z \le 1.3$ (currently in commissioning). All of these ground surveys will produce data with complicated selection effects in both redshift and angular positions due to atmospheric effects. Only a space observatory can observe the brightest emission line (Hα) against a dark sky background, allowing the BAO survey to be done in only a few years with slitless spectroscopy. A slitless spectrograph is a fast and simple implementation that does not require a predecessor full-sky imaging survey to pre-select targets. The JDEM/Omega galaxy redshift survey is valuable for BAO, power spectrum, large-scale structure, and redshift-space distortion measurements, as well as galaxy evolution studies.

**SNe Implementation Specifics:** The SN program requires repeated mapping of small fields near the ecliptic poles at ~5 day intervals to identify type Ia SNe, their redshifts, their apparent brightness over time (light curves), and their reddening (extinction).

All measurements will be performed with the imaging channel using all the filter positions and the R=75 (2-pixel) disperser. A given field will be observed at a fixed spacecraft roll angle for ~90 days, at which time the S/C roll will be changed by ~90° to follow the Sun. The nearly-square monitored field ensures that SNe light curves can be observed continuously. Observations of a given field for periods much longer than the SN duration enables subtraction of the host galaxy flux thus providing an accurate zero-point for each SN brightness measurement. The distribution of SN redshifts observed will depend on the survey strategy chosen. The instrument FOV is large enough to ensure that there will always be SNe present at $z > \sim0.5$, with the upper redshift limit being set by the exposure time per visit. Conversely, at low redshift multiple fields will need to be observed as a consequence of the low cosmic volume, but the exposure time required for each field will be much lower.

The advantages of space vs. ground for high-z ($>\sim z=0.8$) SN surveys are numerous. The low background, compact PSF, high efficiency, and stability of a wide-field imaging space telescope permit rapid monitoring of a large number of $z\sim1$ galaxies, enabling JDEM to find and follow many more supernovae than can readily be discovered from the ground, especially for $z > \sim0.8$. The light curves of SNe can be acquired in a homogeneous, gap-free manner, with superior photometric accuracy, since the space environment has no cloudy or moonlit periods. An absolute photometric accuracy of 1% is more readily maintained in space by monitoring and correcting instrumental effects using celestial sources. While not simple, past missions have demonstrated that this level of accuracy should be achievable by developing an appropriate set of calibration standards, implementing a comprehensive instrument ground calibration program, and if necessary, providing for on-orbit calibration of key parameters like linearity. Spectroscopy in the NIR can produce spectra in the rest-frame visible at redshifts higher than would be possible from the ground, permitting the tracking of spectral features around the time of peak light. Such features can be used to yield lower dispersion in the final Hubble diagram and to test evolution systematics. Finally, photometry in the rest-frame NIR, available to JDEM at low redshift, has also been found to yield intrinsically lower dispersion in the final Hubble diagram, compared to photometry in the rest-frame visible.

**WL Implementation Specifics:** The WL program requires multi-band imaging data to achieve a magni-tude limit of 25.5 at S/N 25. JDEM/Omega can achieve this sensitivity with ~1800 s of integration time per sky field. The total exposure time is accumulated in 3 separate gap-filled passes of 600 s each, with each pass being dedicated to one broadband NIR filter. In combination with ground measurement data, this provides photometric redshifts and tests the wavelength dependency of the shape measurement. Each WL gap-filled pass will be rolled slightly (a few degrees) relative to the other to support the acquisition of BAO data (meeting dispersion direction requirements) as the WL data set is acquired at a sky coverage rate of 3300 deg²/yr.

For JDEM/Omega we have chosen to do WL shape measurements in the NIR with HgCdTe detectors. There are advantages to performing the shape measurements with these detectors in that galaxies are brighter in the NIR relative to the Zodiacal background and galaxy spectra are smoother in the NIR so that wavelength dependent PSF effects are reduced. Integration times are shortened by 25% vs making visible shape measurements in JDEM/DECS. There is however a risk in that HgCdTe detectors have properties, such as interpixel capacitance and persistence, that may affect shape measurement accuracy. That risk is being assessed with laboratory testing and simulation, and will most likely be retired by the end of this calendar year. For this and other reasons (see the risk summary), WL shape measurement is treated as a goal rather than a requirement for JDEM/Omega.

It is important to stress the value of NIR WL color measurements as a complement to ground-based WL shape measurements. The high quality NIR photometry needed to produce photo-z's with uncertainties, biases, and failure sets that satisfy WL requirements, along with the requisite deep NIR spectroscopic training sets, can only be obtained from space, and are provided by JDEM/Omega.

Finally, both shape and photo-z measurements must have systematic errors or biases at least an order of magnitude better than current data. The small and stable PSFs uniquely attainable in space give JDEM/Omega shape measurements the potential to greatly exceed the quantity and quality of ground based data. JDEM/Omega (with orbit and operations designed to emphasize thermal stability), is an excellent platform for resolving galaxy shapes without systematic errors resulting from confusion and atmospheric/gravitational instrumental disturbances/effects.

3. *Of the required measurements, which are the most demanding? Why?*

The key design driver for the overall mission is the need for substantially higher imaging & spectroscopic etendue (large FOV * effective area) than any previous space astronomy mission. This means combining a large, well-corrected optical FOV for each channel with a large complement of low-noise detectors and associated data rate.

Of the three measurement techniques, the requirements for a stable high-quality PSF over a wide FOV make the WL galaxy shape measurements the most demanding overall. The resulting requirements on the optical design, fabrication tolerances, structural stability, and attitude control system performance are more stringent than those imposed by the other techniques. In addition, the combination of fine sampling of the PSF and a wide survey area necessitate a large number of detector pixels.

While not as challenging as the WL measurements, the calibration requirements of the SNe program are worthy of note. Systematic uncertainties in the relative photometric calibrations of each filter bandpass must be kept to under 1%. Systematic uncertainties of even 2% significantly degrade the FoM. Calibrating the large JDEM/Omega focal plane and filter set to this accuracy, and maintaining the calibration over the life of the mission, is a challenge that we have addressed with our calibration plan described in Payload Implementation #9.

The BAO measurements are not as technically challenging as those for either WL or SNe. The optical and ACS performance requirements in particular are significantly relaxed in comparison. The most demanding aspect of the BAO measurements are attaining the faint line flux limit at 2 μm, which necessitates cooling the telescope to below 250 K, and the redshift accuracy, which necessitates careful optical distortion calibrations. Neither of these requirements pushes the state of the art, nor are they driving the JDEM/Omega requirements.

4. *Present the performance requirements (e.g. spatial and spectral resolution, sensitivity, timing accuracy) and their relation to the science measurements.*

See answer 5 for combined response. The performance requirements are given in the "Key JDEM Instrument Design Parameters" box in *Figure 2*.

5. *Present a brief flow down of science goals/requirements and explain why each payload instrument and the associated instrument perfor-*

mance are required.

The performance requirements and their relation to the science measurements are given in the detailed flow down shown in *Figure 2*, starting from our highest level dark energy scientific objectives. These originate from the prior NAS studies, and from the AAAC/DETF, and FoMSWG panels. We show requirements of the three methodologies at their highest-level astrophysical variables. All three methods have statistical requirements on survey size, and have requirements on redshift range and precision, and all have restrictions on the systematic biases. These methodologies define the specific suite of measurements required by JDEM/Omega. All of these correspond to specific sensitivity, wavelength range, filter photometry or spectroscopy measurement needs. The SNe study requires a narrow, deep multi-band photometric survey, the weak lensing study requires a wide area, high-spatial resolution multi-band survey, while the BAO study requires a wide area low-spectral resolution survey. The sensitivity requirements for each study drive the telescope aperture, while performing each technique within one mission drives aperture and field-of-view.

The collection of measurements required for each technique then determines the survey and instrument parameters as shown in the figure. Near Nyquist sampling is achieved with the pixel scale and dithering: the numbers of detectors was selected to achieve a minimum survey rate given an exposure time. With this information the division of instrumentation between imaging and spectroscopy becomes determined and the specific performance requirements are derived from mission optimization and simulation.

The imaging system performs five-band precision photometry for WL and SN studies. Its wide field establishes a large survey rate, ~ 3,300 deg² per year, to deliver a high survey rate for WL and BAO-imaging, and a large number of well-qualified SNe for the supernova measurements. On the filter-wheel, a disperser is provided to establish precise classification of SNe and their host galaxies, and precise redshifts for a significant statistical sample of the WL target galaxies.

Imaging passbands are defined by five filters. These give the SN colors for classification, plus the host galaxy colors and morphology. The photo-z for each field galaxy is determined from its filter-band signature and a galaxy shape is measured for weak lensing science. Radiometric calibration of the entire system

is provided by periodic viewing of white dwarf and solar analog stars.[1]

The weak lensing science faces a tradeoff between better sampling with smaller pixels versus more sky coverage with larger pixels, and one can trade detector costs or FOV against survey duration. We continue to study the sampling trade space for the most cost-effective configurations that accomplish the required survey without systematic errors from aliasing.

The spectrographic system uses a prism for high throughput over 1.1–2.0 μm, and has sufficient resolution ($R_\Theta$=200-240 arc-sec) to meet the BAO requirements. The spectrograph has two channels to provide a counter-dispersed measurement. This eliminates most, if not all, systematic errors associated with the measurement technique.

Two pairs (prime and redundant) of broadband HgCdTe image sensors are located on the focal plane for guiding. The survey fields deliver typically a dozen guide stars to each sensor at any time. The guide stars are 13th to 18th magnitude, permitting star centroid determination to within a few milli-arcsec.[2] This system delivers the observatory stability and knowledge needed during an exposure to enable the WL measurements.

At the observatory level, JDEM/Omega utilizes a Korsch type on-axis three-mirror anastigmat telescope. The 1.5 m aperture rigid light-weight telescope delivers a large, diffraction-limited field of view. Within this field-of-view, ~0.8 deg² are instrumented with ~144 million pixels sensitive to wavelengths from 0.4–2.0 μm. The wide field of view telescope provides an enormous advantage, yielding a high survey rate for a wide, near all sky, WL and BAO survey while also permitting sufficient time-on-target to deliver the requisite sensitivity for the deep SN survey.

6. For each performance requirement, present as quantitatively as possible the sensitivity of your science goals to achieving the requirement. For example, if you fail to meet a key requirement,

---

[1] Bohlin, R.C., Dickinson, M.E., and Calzetti, D., AJ. 122 2118, 2001; Bohlin,R.C. 2002 HST Calibration Workshop p.115; Bohlin, R.C, Riess, A., and de Jong, R."NICMOS count rate dependent nonlinearity in G096 and G141" STSCI ISR-2006-0.

[2] Secroun, A., et al, "A high accuracy small field of view star guider with application to SNAP" Experimental Astronomy v.11, June 2002.

**what will the impact be on achievement of your science objectives?**

In a well optimized mission, parameters that govern each kind of measurement are chosen to yield a satisfactory compromise between alternatives. An adverse impact in any one mission parameter can be partly compensated for by shifting some of the other mission parameters. In *Table 1,* we list JDEM/Omega performance parameters and describe qualitatively the impact of failing to meet each one.

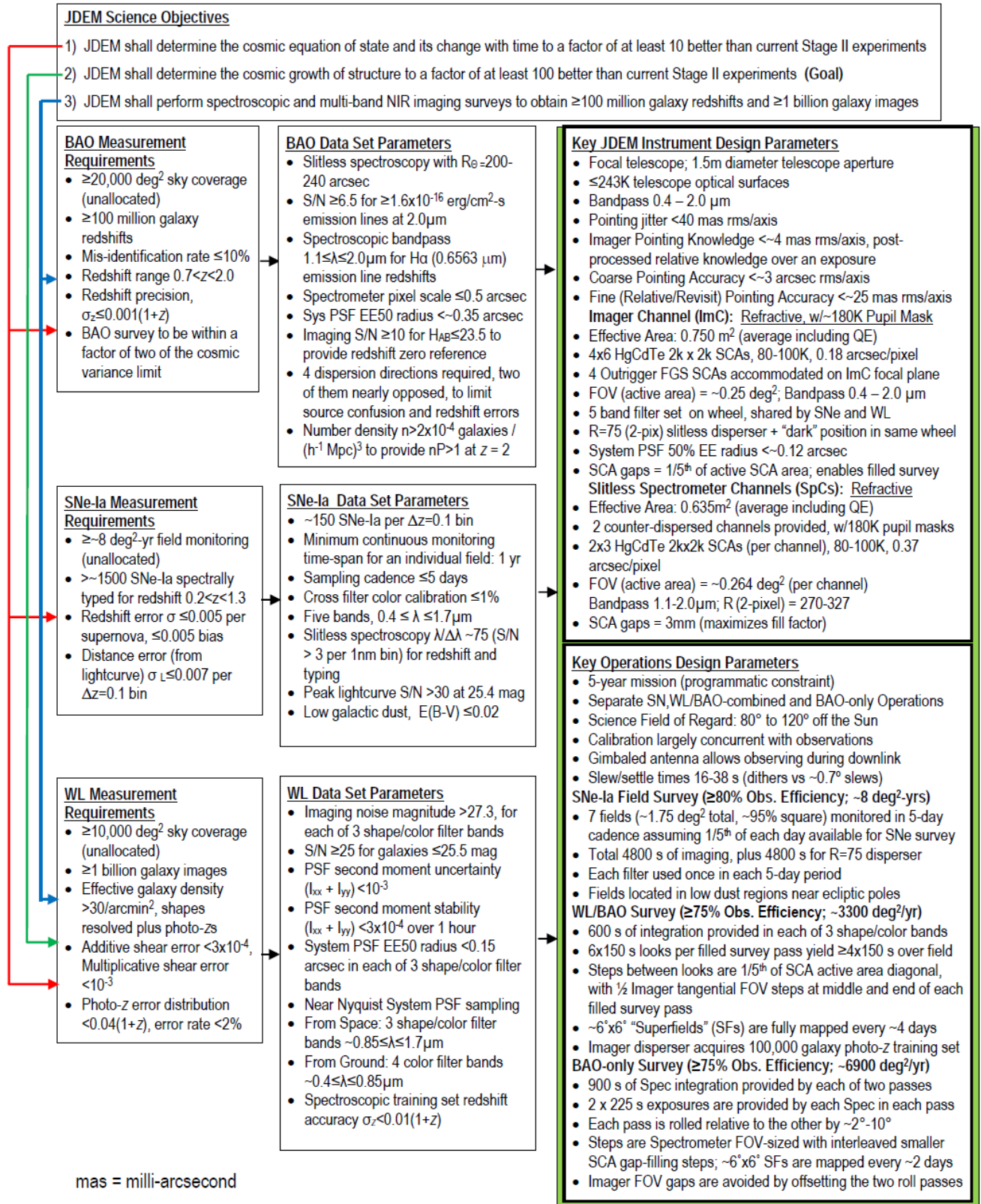

*Figure 2 - Science requirements flow-down. JDEM/Omega has developed end-to-end traceability of its science requirements to instrument parameters.*

*Table 1 - Consequences of failing to meet requirements*

| The Following Requirements Failures: | Will Result in the Following Performance Impacts: |
|---|---|
| 1. Degraded Detector Performance Characteristics (dark current, crosstalk, persistence, linearity / *reciprocity*, charge transfer efficiency degradation) | Increased detector dark current will increase effective noise. Persistence / linearity / reciprocity failure will decrease photometric accuracy or complicate on-orbit calibration procedures. Crosstalk will degrade spatial response purity and affect shape measurements. It will also make precise photometry more difficult. Intra-pixel response nonuniformity will make it difficult to perform accurate photometry and will degrade the WL galaxy shape measurements. *Mitigation*: Laboratory measurements will quantify the risk and will be used to ensure we are within specification. |
| 2. PSF Resolution in Optics | Imaging: Reduced galaxy shear capability; reduced depth (limiting magnitude) for photo-z survey. Spectrometers: Reduced spectral resolving power, reduced redshift accuracy, potential increase in systematic spectral identification errors *Mitigation*: Ground testing will verify the PSF. A 6 DOF adjust mechanism on the secondary mirror will be used for on-orbit sensing and alignment. |
| 3. Degraded Observing Efficiency | Reduced sky coverage per year for WL and BAO Surveys and SNe Field Monitoring result, and will reduce the FoM improvement Sky Survey return achievable during the baseline mission. *Mitigation*: Integrated modeling and a detailed examination of all contributors to observing efficiency has been started and will continue through the build of the Observatory. On orbit, observing strategies can be modified to maximize observing efficiency. |
| 4. Failure to Meet Pointing Knowledge and/or Stability | Pointing stability issues will degrade the effective point spread function, thus decreasing the signal to noise and potentially introducing biases into precision photometry and shape measurements. *Mitigation*: Integrated modeling and a detailed examination of all contributors to observing efficiency has been started and will continue through the build of the Observatory. On orbit, observing strategies can be modified to maximize observing efficiency. |
| 5. Insufficient Stray Light Rejection | In-field stray light: Reduced sky coverage near bright objects *Mitigation*: Tighter polish specifications on optics to minimize scatter. Out-of field stray light: Reduced sky accessibility with reduced sky coverage *Mitigation*: Conventional well baffled system at benign L2 environment. |

## 2. TECHNICAL IMPLEMENTATION

*Payload Instrumentation*

1. Describe the proposed science instrumentation, and briefly state the rationale for its selection. Discuss the specifics of each instrument (Inst #1, Inst #2 etc) and how the instruments are used together.

The JDEM/Omega payload configuration (see *Figure 3* for an optical path block diagram and *Figure 4* for a fields of view layout) is designed to provide the survey data to address all three dark energy observational methods described in the Science section. A 1.5 m aperture focal telescope feeds a single instrument comprised of three observing channels: an Imaging Channel (ImC) covering 0.4 – 2.0 µm and two identical near infrared Spectrometer Channels (SpC) covering 1.1 – 2.0 µm. The instrument uses 2.5 µm long-wavelength cutoff JWST HgCdTe detector material. JDEM requires no detector development. The two SpCs provide the faster survey speeds desirable for a BAO-only survey mode, and because they are dispersed in opposing directions, also provide key source separation information without requiring a later field revisit. The ImC, covering the NIR and optimized to provide good sensitivity down to 0.4 µm in the visible, provides imaging for all three techniques.

The science channels are fed by a Three Mirror Anastigmat (TMA) telescope, which offers a wide field along with a flat focal surface and good correction of low order aberrations. The design uses a focal TMA working at a pupil demagnification of 17.6 (85 mm pupil diameter). A 1.5 meter diameter primary mirror feeds a separate tertiary mirror for the ImC that goes directly to focus at the imaging channel's focal plane, while the two spectroscopic refractive reimaging camera channels are fed afocally via separate 4-mirror telescope collimators. While any one channel can be packaged using a reflective design form, the 3 combined channels can only be packaged for an EELV using refractive spectrometers. In order to separate the beams from the different channels, the TMA is corrected at a large radial field half angle of 0.8 degrees; therefore, the extremes of the different fields of view are separated by up to ~2.5 degrees. The outer barrel assembly and baffles for the primary and secondary mirrors mitigate out of field stray light. The total field of view extent for all three channels is 0.913 deg² (0.764 deg² active area).

The optical telescope assembly (OTA) reflecting surfaces are maintained below 243K to limit the NIR in-band thermal emissions to≤10% of the minimum Zo-diacal background. The instrument volume is also maintained below ~180K to control out-of-band thermal emission. The imaging tertiary is included in the OTA, so the instrument ImC interface (thermal, optical and mechanical) is a real pupil comprised of a pupil mask and filter wheel. The spectrometer collimators are included in the OTA, so the interface to each spectrometer is a collimated beam, allowing easy spectrometer interface testing prior to payload integration.

The secondary mirror has a 6 degree of freedom mechanism to adjust focus and alignment. The primary, secondary and tertiary mirrors are made from Zerodur. Each collimator feed consists of two Zerodur mirrors followed by two CaF$_2$ refractive corrector plates. The OTA structure is manufactured from composites to minimize mass and thermal distortions while providing adequate stiffness.

The use of CMOS-multiplexer (readout integrated circuit) based hybrids with non-destructive readouts, supporting noise reduction, and electronic shuttering eliminates the need for a shutter mechanism. Sample up the Ramp processing is used during all observations with intermediate read-outs at an ~1.3 sec frequency being combined to produce one image for each observation (Offenberg et al., PASP, 117, 94, 2005). All detectors in the three instrument channels are identical, simplifying detector production and sparing. The baseline design is a direct reuse of 2K x 2K JWST HgCdTe detectors with a 2.5 µm long-wavelength cutoff and 18 µm pixels operating at 80 to 100K.

An ImC filter wheel provides 5 filters, a blank, and an R-75 (2-pixel) dispersing element for executing the SNe and WL programs, including the WL photo-z training set.

The imager performs the critical WL shape measurements. Its system error budget and resulting PSF include diffraction, visible-quality polished optics, well-controlled pixel cross-talk effects and 40 mas/axis RMS jitter. These combine to form a system wavefront error of 125 nm RMS, for a working diffraction limit of λ₀=1.67 µm and a system PSF EE50 radius of ~0.12 arcsec.

Though not formally a part of the scientific instrumentation, there are additional imaging detectors used for fine guidance. During normal imaging operation, the four "outrigger" detectors located at the ends of the ImC are used for guiding. When the disperser is inserted by the filter wheel, these outriggers will no longer see undispersed stellar images, so a separate on-axis Fine Guidance Sensor provides this function. Further discussion of the FGS implementation is in the spacecraft section.

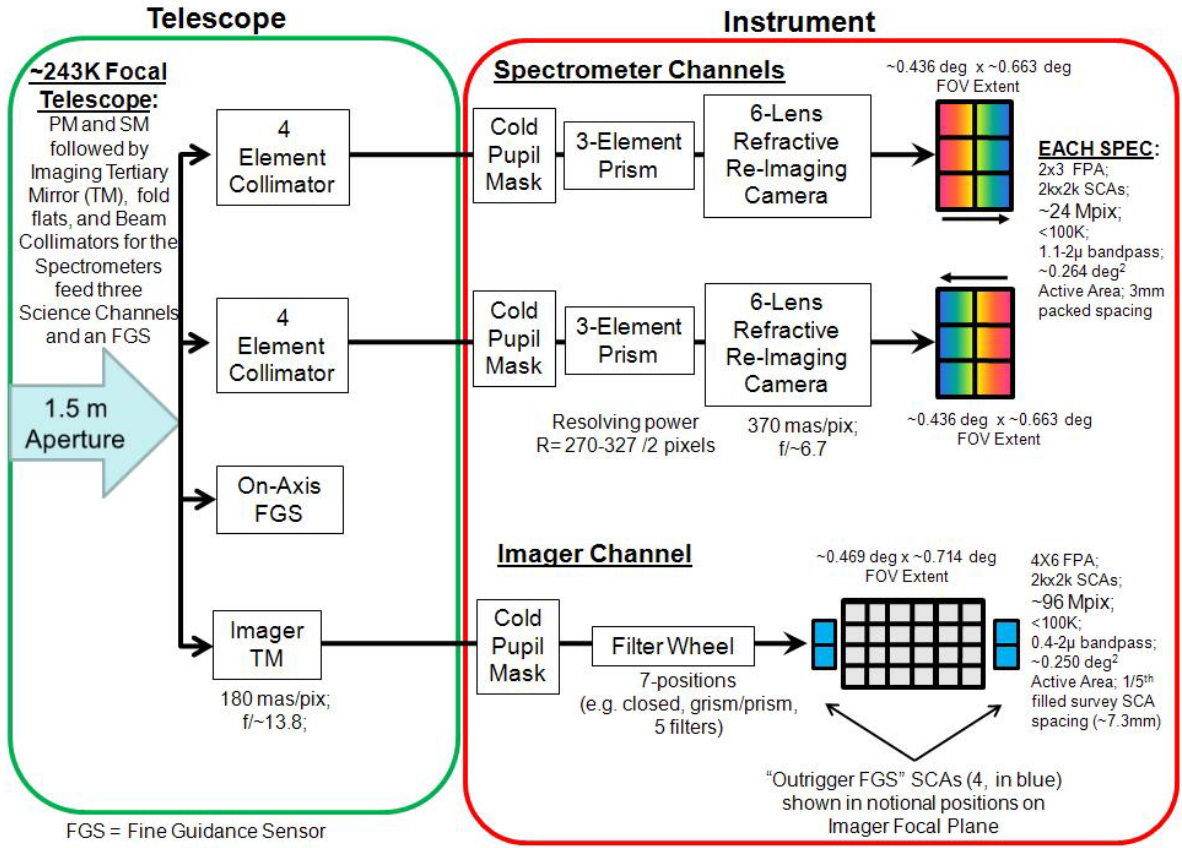

*Figure 3 – Payload Block Diagram (*ImC *SCAs not to scale with* SpC *SCAs)*

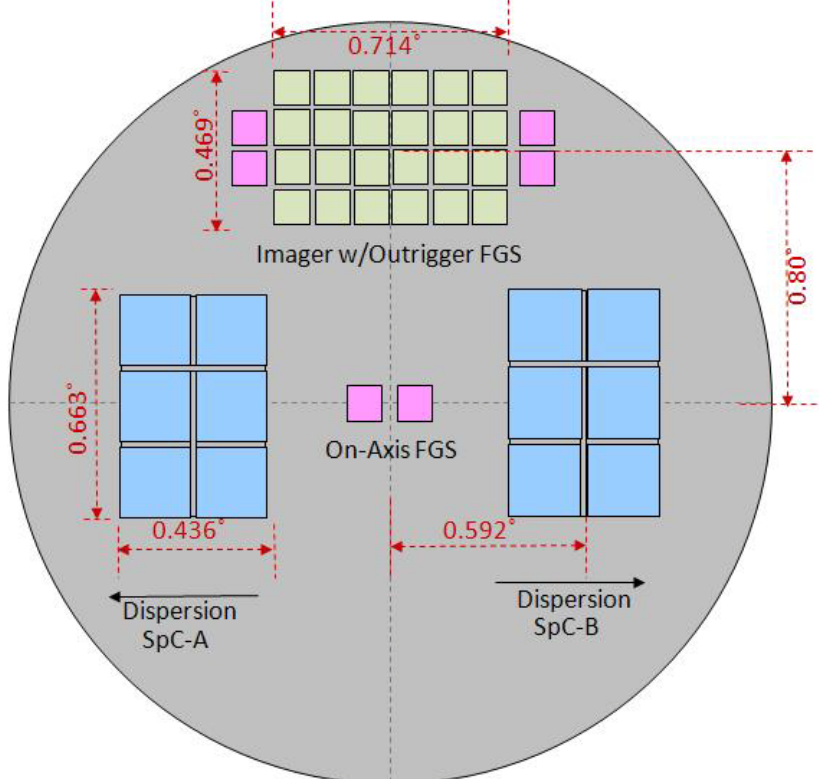

*Figure 4 – Payload FOVs at Telescope Intermediate Focus (*ImC *SCAs not to scale with* SpC *SCAs)*

The key drivers for the selection of the hardware were the need to achieve the required FoM within 5 years, the need to minimize the risk related to acquiring the WL galaxy shapes to the required accuracy, and the need to minimize the risk related to achieving mission success. The FOM and mission life requirements drove the decision to pursue three techniques and drove the telescope size and FOV (to achieve an acceptable etendue, the product of effective area and FOV), while the WL shape requirements drove the selection of a focal telescope that would simplify the ImC's optical path. Dual spectrometers were chosen in order to obtain the required field of view to enable a linked WL/BAO survey. Two spectrometers also enable opposed dispersions which minimizes systematic errors and eliminates the need to revisit the field several months after the initial visit.

JDEM/Omega observes in a survey mode to acquire WL and BAO data, with all its channels operating simultaneously and integrating synchronously in step with the spacecraft pointings that implement the sky coverage strategy. The ImC and SpC fields-of-view are carefully arranged in both angular extent and rotation, and the SCAs in the ImC are displaced relative to each other by 0.2 of the SCA's active area to enable the sky survey scheme described in the response to question 1 in the Mission Design section. JDEM/Omega executes a field monitoring strategy to enable the SNe technique, revisiting fields on a 5-day cadence to detect SNe, track light curves, identify Ia types, measure redshift, determine absolute brightness, and estimate reddening (i.e. extinction). Every ~90 days the observatory must rotate ~90 degrees in order to keep the Sun from striking the cold side of the Observatory, and the SNe fields are nearly square so they can be rotated and still monitor the same sky fields.

In summary, the BAO measurements use the two spectrometer channels and imaging in any filter; the WL measurements use the imaging channel with 3 NIR filters for shapes and photo-z's, and the disperser for photo-z calibrations; and SNe measurements use the imaging channel with all filters and the disperser.

2. Indicate the technical maturity level of the major elements and the specific instrument TRL of the proposed instrumentation (for each specific Inst #1, Inst#2 etc), along with the rationale for the assessment (i.e. examples of flight heritage, existence of breadboards, prototypes, mass and power comparisons to existing units, etc). For any instrument rated at a Technology Readiness Level (TRL) of 5 or less, please describe the rationale for the TRL rating, including the description of analysis or hardware development activities to date, and its associated technology maturation plan.

The JDEM/Omega design is optimized to use mature technology for space flight. All of the components of the Payload Instrumentation are at TRL 6 level or higher and are based upon flight heritage. See *Table 3* for the flight heritage and TRL assessment.

3. In the area of instrumentation, what are the three primary technical issues or risks?

See *Table 13* in the Programmatics & Schedule section. Risk numbers 3, 4 and 5 are the three primary technical risks for the instrumentation.

4. Fill in entries in the Instrument Table. Provide a separate table for each Instrument (Inst #1, Inst #2 etc). As an example, a telescope could have four instruments that comprise a payload: a telescope assembly, a NIR instrument, a spectrometer and a visible instrument each having their own focal plane arrays.

See *Table 4* and *Table 5*.

5. If you have allocated contingency please include as indicated along with the rationale for the number chosen. If contingency is unknown, use 30% contingency.

30% contingency is used for both mass and power.

6. Fill in the Payload table. All of the detailed instrument mass and power entries should be summarized and indicated as Total Payload Mass and Power as shown in the table

See *Table 6*.

7. Provide for each instrument what organization is responsible for the instrument and details of their past experience with similar instruments.

The mission is currently in pre-Phase A and is progressing with the definition of the reference mission design and assignment of hardware roles. An MOU is in place between DOE and NASA. Roles and responsibilities are being determined between the agencies.

8. For the science instrumentation, describe any concept, feasibility, or definition studies already performed (to respond you may provide copies of concept study reports, technology implementation plans, etc).

Concept studies led by UCB, JHU, NOAO, and NASA are summarized in Section 7.0 in *Table 18.* The ADEPT, DESTINY and SNAP concepts were all reviewed by BEPAC. While there is some variation in the implementation depending upon the emphasis of each team, all enable more than one dark energy technique and are similar in terms of the type and scale of instrumentation. All rely on adapting existing technology rather than new technology development.

9. For instrument operations, provide a functional description of operational modes, and ground and on-orbit calibration schemes. This can be documented in Mission and Operations Section. Describe the level of complexity associated with analyzing the data to achieve the scientific objectives of the investigation. Describe the types of data (e.g. bits, images) and provide an estimate of the total data volume returned.

The Operational Modes for the JDEM/Omega Instrument that are used throughout the mission are Observing Modes, Engineering Modes, and Safe Modes.

A single Instrument Observing mode is required to implement virtually all of the sky tiling (WL and BAO) and Field Monitoring (SNe) activities, easing I&T verification efforts and limiting failure modes. This is a series of exposures, each separated by a slew to a new pointing position followed by an FGS-controlled settle. One image is produced for every exposure via the use of Sample up the Ramp processing and sent to the ICDH for square root and lossless compression. The filter wheel's position can optionally be changed between the exposures. The slew could be star tracker controlled, gyro controlled, or could be an offset relative to the FGS pointing. The data volume for the observing mode drives the data rate for the mission. Daily data volume is addressed in the Mission Operations section.

Engineering Modes are provided to configure the Instrument, diagnose/prevent Instrument problems, verify Instrument performance, and perform Instrument maintenance. Examples of some Engineering Mode activities would be Contamination Prevention (cooldown control) Heater Mode, Contamination Removal Heater Mode (heat sensitive portions of the Instrument to remove contamination buildups) and Diagnostic Mode (used in commissioning and to allow verification of instrument software processing).

In the Safe Mode, the Instrument is completely powered off. Survival heaters are provided for the instrument boxes mounted on the spacecraft and on the Instrument Cold Sensing Assembly. In order to maintain the thermal stability required by the observing sensors this mode would only be used during launch and if maximum load-shedding were required. Though it would be preferable to be able to enter safe mode in a set configuration, the goal is to not require any Safe Mode Entry warning. In non-power critical safing events, the instrument would only be partially powered down to minimize the time to return to operations.

Past experience with space imaging and spectroscopic missions leads to the conclusion that JDEM/Omega has stringent calibration requirements in a number of areas. The general JDEM/Omega strategy is to use ground calibration methods to the extent possible, reserving on-orbit calibration to verification of the ground results and extending the calibrations where ground calibration may not be effective. To maintain the calibration requirements over the entire mission, not only are the calibrations important, but so are estimates of calibration stability. The latter will determine the need for and frequency of on-orbit calibrations. The JDEM/Omega calibration program will place strong emphasis not only on the areas requiring calibration, but also on the verification of these calibrations, either on the ground or in orbit, using multiple techniques as cross-checks. The SN fields are observed repeatedly over the lifetime of the mission, providing excellent opportunities to develop and use sky calibration standards.

All optical and detector components will be calibrated at the component, subsystem and instrument levels. These data will be used to feed an integrated instrument calibration model that will be verified using an end-to-end payload-level thermal vacuum test. This test will involve a full-aperture (1.5 meter) diameter collimated beam that will test for optical wavefront error as well as photometry.

The three observational methods have different calibration demands on instrument parameters and their accuracy. The SN Survey places the most stringent demands on photometric calibration. White Dwarfs and other suitable sky calibration targets will be used to calibrate the linearity of the imager over several orders of magnitude. This linearity will be tested on the ground, and verified with an on-orbit relative flux calibration system, if necessary. It will also be necessary to understand the intra-pixel response function (quantum efficiency variations within a pixel), which will be fully characterized by ground testing for each detector.

For the WL Survey, the requirement for galaxy ellipticity accuracy places significant demands on both the optical and detector subsystems. The uniformity and stability of the point spread function (PSF) needs to be strictly controlled and monitored to ensure a successful mission. This drives the need to characterize the intra-pixel response and the inter-pixel response (capacitive cross-coupling with nearest neighbors) for magnitude as well as spatial and temporal variations. It is likely that the combined PSF effects will have some variability on time scales of a single exposure. These residual effects will be continuously monitored with the observatory attitude control system and field stars.

The BAO survey relies primarily on the spectrometer channels, which are not driving the calibration requirements for the mission. Established calibration techniques used for other space missions should be adequate to meet the relatively loose photometric and morphological requirements. The larger plate scale in the spectrometers may demand some attention to the spatial effects such as intra-pixel response, but not to the degree required by the WL Survey.

A description of the ground processing of the science data is provided in the response to Mission Operations question 4.

10. Describe the instrument flight software, including an estimate of the number of lines of code.

The Science Instrument FSW will be resident in the Instrument Control Electronics (ICE) box running on a RAD-750 processor. The proposed hardware architecture is similar to that chosen for the Lunar Reconnaissance Orbiter (LRO) mission. The NIR Instrument FSW's specific responsibilities include initialization, control, and readout of detector electronics; science data compression and packetization; communication with the spacecraft bus; instrument mode management and execution; instrument mechanism management; active thermal control of focal plane electronics; and instrument health and safety monitoring

On-board data processing will be relegated to hardware (ASIC & FPGA), so Instrument FSW will not be directly involved. However, it will be responsible for initialization, control, and readout of said hardware.

Once data is read out into the ICE box (where FSW resides), FSW will initiate a 2:1 compression on data which again is performed in hardware. FSW will then packetize data into the CCSDS format and then transmit data to the spacecraft bus using the SpaceWire interface. The SpaceWire interface is also implemented in hardware.

Instrument modes managed by FSW include boot/initialization Mode, science mode, commissioning mode (includes possible calibration and diagnostic sub-modes), and survival mode/safe mode

Instrument FSW will be responsible for command and control of a 7-position filter wheel. This mechanism has a sensor specifying the current orientation/configuration of the mechanism, and it is the responsibility of the Instrument FSW to read that data from the mechanism electronics and process the raw data into engineering units.

Active thermal control of the Focal Plane Electronics (FPE) will be managed by Instrument FSW. Processing required to perform this function requires monitoring temperatures and simple commanding of heaters via heater control hardware.

The Instrument FSW will support diagnostic functions for detecting and troubleshooting potential instrument health and safety problems. The operational philosophy for instrument Fault Detection and Correction (FDC) capabilities is fail-safe. In the event of an in-flight anomaly, the science instrument will fail-safe to Safe Mode rather than fail-operational, and any switching to redundant components will be ground-commanded rather than autonomous.

Finally, the Instrument FSW will be composed of a Core Flight System (CFS) which is platform-independent, mission-independent FSW code developed and maintained by GSFC. Although some new code needs developing for this project, a reusable portion of the total code needs no additional development. That reusable portion is currently flying successfully on LRO. An estimate of the lines of code and amount reusable from previous missions is shown in *Table 2*.

11. Describe any instrumentation or science implementation that requires non US participation for mission success.

No foreign participation is required. All necessary scientific and technical personnel, knowledge, capabilities, technology, facilities and infrastructure reside within NASA, U.S. educational institutions and industry.

12. Please provide a detailed Master Equipment List (MEL) for the payload sub-categorized by each specific instrument indicating mass and power of each component. This table will not be counted in the page totals.

The Payload MELs are included in Appendix A in the restricted data submission.

13. Describe the flight heritage of the instruments and its subsystems. Indicate items that are to be developed, as well as any existing hardware or design/flight heritage. Discuss the steps needed for space qualification.

The flight heritage for the payload subsystems and components is presented in *Table 3*. All subsystems draw on NASA/GSFC's extensive experience in building instruments for space missions and predominantly are at TRL ≥6. The cold lens optical mounts and large focal planes are items that may pose technical challenges. We are currently building and testing Engineering Development Units to retire this risk. The details of the state of these technologies and the risk reduction efforts are in the Enabling Technologies section and *Table 13*, respectively.

*Table 2 - Instrument SLOC Estimate*

| JDEM Instrument FSW Modules | SLOC | Reused % | Reused SLOC | New SLOC | Heritage |
|---|---|---|---|---|---|
| core Flight Executive (cFE) | 19,600 | 100% | 19,600 | 0 | LRO - GSFC Heritage SW |
| Core Flight System (CFS) | 15,700 | 100% | 15,700 | 0 | GSFC Heritage SW |
| Memory Scrub | 1,700 | 100% | 1,700 | 0 | LRO Heritage |
| cFE/CFS Mission Config. Param | 800 | 0% | 0 | 800 | New for JDEM Instrument |
| 1553 RT Task | 1,000 | 75% | 750 | 250 | SDO RT Heritage |
| SpaceWire Task | 3,000 | 75% | 2,250 | 750 | LRO Heritage |
| Fault Detection & Correction | 500 | 0% | 0 | 500 | New for JDEM Instrument |
| Instrument Management | 10,000 | 0% | 0 | 10,000 | New for JDEM Instrument |
| JDEM Instrument Estimate | 52,300 | | 40,000 | 12,300 | |

Assumptions:

Instrument processor is a RAD 750.
cFE/CFS applications (re-usable software) are used

The cFE is a set of services and an operating environment to support and host flight software applications. Based on the core infrastructure and API, reuse library components and new applications can be put together to easily create new systems. cFE includes the following applications: Software Bus, Event Handler, Time Management, Table Management, Executive and Task Services.

The Core Flight System is a platform-independent, mission-independent Flight Software environment composed of a reusable core flight executive (cFE), selected cFE-compliant Applications, and an Integrated Development Environment (IDE). CFS includes the following applications: Stored Command, File Manager, Scheduler, Limit Checker, Checksum, Housekeeping, Memory Dwell, Memory Manager, Data Storage, Health and Safety.

Table 3 – Payload Heritage

| Instrument Subsystem | | Flight/Test Heritage | Existing Hardware | Items to be Developed | TRL | Steps Needed for Space Qualification |
|---|---|---|---|---|---|---|
| Optics | TMA Telescope | GeoEye | None | Primary & Secondary Mirrors, Imaging Channel Feed, Spectrometer Channel feed, | 7 | Build & test prototype components |
| | NIR | NIRCam, Spitzer/IRAC, LDCM, WFC3, NIRSpec, Cassini/CIRS | None | Imaging Channel, Spectrometer Channel collimator | 6 | Build EDU science instrument channels Environmentally qualify EDU(s): sine & random vibration, acoustic, shock, static pull/sine burst, thermal vacuum cycling |
| | FGS | HST, JWST | None | Camera Optics | 6 | |
| | Filters & SN disperser | HST, Sptizer, JWST | None | Imaging filters | 6 | |
| Optical Mounts | Mirrors | JWST, Spitzer/IRAC Spitzer/OTA | None | Prototype, EDU, and Flight | 6 | Complete lens mount risk reduction effort that includes environmental qualification: sine & random vibration, acoustic, shock, sine burst, proof test, thermal vacuum cycling |
| | Lenses | JWST/NIRCam, Spitzer/IRAC LDCM/TIRS, Cassini/CIRS | None | Prototype, EDU, and Flight | 6 | |
| Detector | HgCdTe hybrid array | JWST/NIRSpec | Teledyne H2RG | Flex/ribbon cable | 7 | Develop Engineering Development IR Focal Plane Assembly. Environmentally qualify EDU FPA. Demonstrate detector and front end electronics performance Peer reviews |
| | HgCdTe Front End Elec. | HST ACS Repair, JWST | Teledyne SIDECAR | Package & PWB | 7 | |
| Mechanisms | Filter Wheel | HST/ACS, HST/WFC3, TIRS, JWST/OSIM, IRMOS | None | Brackets, wheel, hub, shaft, motor, filter mounts | 9 | Prototype Development Unit EDU subject to complete environmental qualification program to include random, sine, sine burst & shock where applicable, thermal vacuum cycling, and |
| | Tel. Cover | Orbital Express, Falconsat, Kepler | None | Aperture cover dome, bolt catchers, brackets, kick-off spring & snubber assem- | 8 | |

| Instrument Subsystem | | Flight/Test Heritage | Existing Hardware | Items to be Developed | TRL | Steps Needed for Space Qualification |
|---|---|---|---|---|---|---|
| | | | | blies, release mechanism | | EMI/EMC tests. |
| | Tel. Secondary Mirror | JWST, LISA, SPOT, TIRS | Nexline actuators, BEI linear encoder | flexures, structure, mirror mount | 6 | After environmental qualification, EDU High Duty Cycle Mechanisms will be subjected to 2X life test. |
| Structure | precision composite structure & optical bench | HST WFC3 Optical Bench, ACTS Truss, Swift optical bench, SDO Optical Bench, LRO Instrument Support structure, JWST/ISIM | None | Optical Bench, Outer Barrel Assembly and supports, Aft Metering Structure and Secondary Support Structure | 9 | Develop finite element model and analyze with NASTRAN Peer review & CDR Fabricate & integrate structure Qualify structure with component mass models: mass properties, modal survey, sine & random vibration, acoustic, shock, static pull/sine burst, thermal cycling |
| Thermal | Z93 White Paint for Radiators | AIM, CALIPSO | Commercially available | None | 9 | Develop thermal model based on instrument temperature limits and thermal loads. Establish component/part specifications. Peer review Integrate components/parts. Thermal vacuum test @ component, instrument, & observatory levels. Correlate thermal model with test results |
| | MLI with germanium black Kapton outer layer | ST5, SDO,LRO | Built in-house | None | 9 | |
| | Kapton film heaters | Swift; WMAP,LRO | Commercially available | None | 9 | |
| | Flexible Heat Straps - AL 1100 Stacked Aluminum Foils | WMAP, JWST/ISIM, LDCM\TIRS | Commercially available | None | 9 | |
| | AL 1100 Radiator facesheet operating at ~75K | WMAP | Commercially available | None | 9 | |
| | Mechanical Thermostats | Swift, SDO, STEREO, LRO | Commercially available | None | 9 | |
| | Electronic Heater Controllers | Swift; TRMM | None | Replicate design | 9 | |
| | Interface Filler Materials (e.g., Nusil, CHOTHERM) | ST5, WMAP,SDO, LRO | Commercially available | None | 9 | |
| | Gamma-alumina thermal isola- | ST5, WMAP, GPB | None | Fabricate isolators | 9 | |

| Instrument Subsystem | | Flight/Test Heritage | Existing Hardware | Items to be Developed | TRL | Steps Needed for Space Qualification |
|---|---|---|---|---|---|---|
| | tors | | | | | |
| | Thermistors/PRT's | ST5, Swift; WMAP, SDO, LRO | Commercially available | None | 9 | |
| Electronics | Payload Flight Single Board Computer – | LRO, SDO, GLAST, JWST, Deep Impact, AEHF, STEREO, et al. | BAE Rad750 | None | 9 | Develop board designs Establish parts requirements & spec's. Worst case, stress, & radiation effects analyses Peer review Build hardware Environmentally test: EMI, dynamics, and thermal vacuum at box, system, instrument, and observatory |
| | On-board signal processing and supercomputing resources | H/W LRO JWST,MMS | FPGA Actel RTAX2000 | Algorithms | H/W 9 Algo-rithm 7 | |
| | Instrument High speed multi-channel data Payload Compression Modules | JWST, MMS | EDU/GSFC Pen Shu | Under Production by Aero-flex | 7 | |
| | Random Access Volatile Memories | LRO, SDO, HSt, JWST, Kepler | External Syn-chronous Dynamic RAM (SDRAM) 125MB Modules Maxwell, Aeroflex, Honeywell | need to be qualified for radi-ation latchup using existing facilities | 6-9 | |
| | • PDU Components<br>• Solid State Relays<br>• Radhard Analog Multip-lexers<br>• Operational Amplifiers<br>• Radhard Analog –to-Digital Converters(ADC)<br>• Low Voltage Differential (LVDS) Interface Drivers<br>• Space Wire Interface | LRO, JWST, SDO, MMS, HST | Commercially available Aeroflex, Ho-neywell, et al. | None | 9 | |

| Instrument Subsystem | | Flight/Test Heritage | Existing Hardware | Items to be Developed | TRL | Steps Needed for Space Qualification |
|---|---|---|---|---|---|---|
| | (SpW) Drivers. | | | | | |
| | • 1553 Interface Drivers | | | | | |
| | Cable & Harnesses | HST, JWST | None | Built in-house-needs to be qualified | 6 | |
| | Chassis | | None | Built in-house Multiple large chassis (9 Boxes) & grounding | 6 | |
| Flight Software | cFE (Reusable code) | LRO, 582 Heritage SW | 19,600 SLOC | None | 9 | Identify FSW requirements & HW/SW interfaces Design FSW and identify data flow Develop modules Unit test modules Code Walkthroughs FSW SRR, PDR, CDR Build test of the integrated modules on FSW Testbed System test of the integrated modules on high fidelity FSW testbed and Flatsat Test on flight hardware system during environmental tests (EMI and Thermal Vac) at instrument and Observatory levels |
| | CFS (Reusable code) | 582 Heritage SW | 15,700 SLOC | None | 8 | |
| | cFE/CFS Mission Config Param | None | None | 800 SLOC | 6 | |
| | Memory Scrub | LRO | 1700 SLOC | None | 8 | |
| | 1553 RT code | SDO RT, ELC | 750 SLOC | 250 SLOC | 7 | |
| | SpaceWire Task | LRO | 2,250 SLOC | 750 SLOC | 7 | |
| | Fault Detection & Correction | None | None | 500 SLOC | 6 | |
| | Instrument Manager | JWST NIR Spec (SW Design Heritage) | None | 10,000 SLOC | 6 | |

*Table 4 - Instrument Table for the JDEM/Omega Instrument*

| Item | Value | Units |
|---|---|---|
| Type of instrument | Multi-channel NIR/VIS Imager and NIR Spectrometer | |
| Number of channels | 3 channels<br>2 Identical SpCs: 1.1 $\mu$m – 2.0 $\mu$m<br>1 ImC: 0.4 $\mu$m – 2.0 $\mu$m | |
| Size/dimensions (for each instrument) | 1.5 x 1.1 x 0.95 | m x m x m |
| Instrument mass **without** contingency (CBE*) | 218 | Kg |
| Instrument mass contingency | 30 | % |
| Instrument mass **with** contingency (CBE+Reserve) | 283 | Kg |
| Instrument average payload power **without** contingency | 337 | W |
| Instrument average payload power contingency | 30 | % |
| Instrument average payload power **with** contingency | 438 | W |
| Instrument average science data rate^ **without** contingency | 2.15 | Gbps |
| Instrument average science data^ rate contingency | 50 | % |
| Instrument average science data^ rate **with** contingency | 3.22 | Gbps |
| Instrument Fields of View (extents, including sensor gaps) | NIR Spec A: 0.436 x 0.663<br>NIR Spec B: 0.436 x 0.663<br>Imager: 0.469 x 0.714 | degrees |
| Instrument Fields of View (active areas) | NIR Spec A: 0.264<br>NIR Spec B: 0.264<br>Imager: 0.250 | degrees$^2$ |
| Pointing requirements (knowledge) | ≤4 (**Imager** only, post-processed knowledge of relative motion during a WL shape integration) | milli-arcseconds RMS per axis |
| Pointing requirements (control) | Coarse Pointing (star tracker): ≤3000<br>Fine Pointing (Relative/Revisit using FGS): ≤ 25 | milli-arcseconds RMS per axis |
| Pointing requirements (stability) | 40 (per integration time) | milli-arcseconds RMS per axis |

*CBE = Current Best Estimate.
^ Science Data Rate is the direct digitization data rate of the sensors before on-board Sample up the Ramp processing or any lossy or lossless compression.

Table 5 - Instrument Table for the JDEM/Omega Telescope

| Item | Value | Units |
|------|-------|-------|
| Type of instrument | 3 Mirror Anastigmat Telescope | |
| Number of channels | 3 tertiary mirrors in combination with 2 collimators feed a focal beam to the ImC and collimated beams to the 2 SpCs<br>1 additional pick off mirror for the On-Axis FGS | |
| Size/dimensions (for each instrument) | Primary mirror diameter is 1.5 m<br><br>Distance between primary to secondary mirror is 2.15 m<br><br>Outer barrel is 2.9 m long with a diameter of 1.8 m. | m |
| Instrument mass **without** contingency (CBE*) | 789 | Kg |
| Instrument mass contingency | 30 | % |
| Instrument mass **with** contingency (CBE+Reserve) | 1026 | Kg |
| Instrument average payload power **without** contingency | 115 (incl. heaters) | W |
| Instrument average payload power contingency | 30 | % |
| Instrument average payload power **with** contingency | 150 | W |
| Instrument average science data rate^ **without** contingency | N/A | kbps |
| Instrument average science data^ rate contingency | N/A | % |
| Instrument average science data^ rate **with** contingency | N/A | kbps |
| Instrument Fields of View (if appropriate) | N/A | degrees |
| Pointing requirements (knowledge) | N/A | degrees |
| Pointing requirements (control) | N/A | degrees |
| Pointing requirements (stability) | N/A | deg/sec |

*CBE = Current Best Estimate.
^Instrument data rate defined as science data rate prior to on-board processing

Table 6 - Payload Mass and Power Table

| Payload Element | Mass Current Best Estimate (CBE) (kg) | Mass Contingency (%) | Mass CBE Plus Contingency (kg) | Power Current Best Estimate (CBE) (W) | Power Contingency (%) | Power CBE Plus Contingency (W) |
|------|------|------|------|------|------|------|
| Instrument | 218 | 30 | 283 | 337 | 30 | 438 |
| Telescope | 789 | 30 | 1026 | 115 | 30 | 150 |
| Total Payload | 1007 | 30 | 1309 | 452 | 30 | 588 |

## Mission Design

1. Provide a brief descriptive overview of the mission design (launch, launch vehicle, orbit, pointing strategy) and how it achieves the science requirements (e.g. if you need to cover the entire sky, how is it achieved?).

JDEM/Omega will be launched from Cape Canaveral, Florida aboard an Evolved Expendable Launch Vehicle (EELV) that will place the observatory into a transfer trajectory to an Earth-Sun L2 libration point orbit. The on-board propulsion system will be used to perform mid-course adjustments, orbit maintenance and momentum dumps. Orbit maintenance and momentum dumps are managed to coincide with attitude maneuvers so as to avoid any significant impact on observing efficiency. The L2 libration point orbit has been selected to provide the thermal stability, minimum stray light and large sky coverage needed to make the required science observations. A mission life of 5 years is specified with the opportunity for an extended mission of up to an additional 5 years.

SNe observations require regular monitoring of small (e.g. few deg$^2$) fields over extended periods of time (>1 year) while the WL and BAO surveys require mapping large sky areas as rapidly as exposure times permit. Combining these sky coverage requirements with stray light and solar array exposure considerations, the observatory is designed to have a Field of Regard (FOR) between 80 and 120 degrees from the sun with no azimuthal constraint about the sun line. During SNe observations, the observatory repeatedly monitors ~2 degree$^2$ of sky area using approximately square fields within 10° of the ecliptic pole(s) with a 5 day cadence. To enable continuous monitoring, the observatory roll angle is inertially fixed for ~90 day periods, then rotated ~90 degrees within the square field to keep the sun angle within 45 degrees of the maximum-power roll angle while maintaining field coverage. The area covered by the WL and/or BAO surveys is broken into a series of ~6° x ~6° "SuperField" (SF) observations, each comprised of a programmed sequence of small slews/dithers over an ~4 day (WL/BAO-combined) or ~2 day (BAO-only) period. The SF locations chosen ultimately stitch together into a contiguous map, and are selected within the FOR according to a schedule that accounts for Zodiacal brightness, the Galactic plane, and other geometric/thermal constraints. Given the L2 vantage point and the 5 year mission life, the scheduling constraints to achieve the WL and BAO sky coverage are not challenging.

The JDEM/Omega design provides the flexibility to support different observing strategies. Table 7 gives the sky coverage and SNe detection rates per dedicated year. The observing time can be divided in different ways to provide large sky surveys and more than a thousand SNe detections. One example that produces a DETF FoM of ~950 is based on a 2 year WL/BAO combined survey, a 2 year BAO only survey and 1 year SNe survey. The JDEM science team will work with the Project to determine the optimal combination of observing strategies.

| Observing Strategy | Return |
|---|---|
| WL/BAO combined | 3,300 deg$^2$/yr |
| BAO only | 6,900 deg$^2$/yr |
| SNe | >1500 SNe |

**Table 7 - *JDEM/Omega Observing Capabilities***

The WL/BAO combined observing mode reduces chromatic dependencies in the WL measurements by implementing three separate passes, each with ≥4 random dithers and a different filter, over each SF. This WL/BAO combined observing strategy also provides a deep BAO spectroscopic survey, which is critical for limiting observational systematic uncertainties and measuring redshift space distortions.

The BAO-only survey will be conducted at about twice the speed of the WL/BAO combined survey due to the total field of view of the spectrometers being roughly twice that of the imager and the required integration times being similar. The speed of the survey is enhanced through the use of slitless spectroscopy, with four views (two opposite dispersions at two different roll angles) being acquired to control losses due to source confusion.

SNe fields are located near the ecliptic poles and subfields are visited on a five day cadence for SNe light curve tracking. Imaging observations in multiple filters over the bandwidth from 0.4 to 2.0 µm and spectrometry via the disperser in the filter wheel are performed during each subfield observation. The amount of precision required for these imaging dithers (~25 milli arc seconds, rms) is achieved under the attitude control capability specified in *Table 11* of the spacecraft section.

2. Describe all mission software development, ground station development and any science development required during Phases B and C/D.

Mission software development encompasses ground system and flight software. The ground system is comprised of 4 major elements: the Mission Operations Center (MOC); the Science Operations Center (SOC); the Science Support Center (SSC) and a Data Archive. [See the response to Question #5 in the Mission Operations section for a description of each element.] Each of the elements will be developed separately. The MOC software is based on heritage command and telemetry systems (e.g. ITOS, ASSIST, ECLIPSE) as well as GOTS and COTS products for planning and scheduling, trending, and paging. A heritage system will be baselined and then the development contractor will add mission unique capabilities to satisfy requirements that are not included in the heritage system. For the SSC, a heritage planning and scheduling system (such as TAKO used on Fermi) will modified for JDEM/Omega mission unique functions such as dithering and conducting revisits to SN targets. The SOC software development effort includes integrating science algorithms provided by the JDEM science teams into a pipeline processing system. Pipeline processors used on HST can be modified for JDEM/Omega. The data archive will most likely be built using an existing facility and archive code, with some JDEM/Omega unique updates (e.g. search by, etc.). Each system will produce multiple builds and releases, providing an incremental approach to software development. After each build, there will be independent testing of the system followed with testing by the respective operations teams. This is followed by interface testing between the various elements. This is repeated after each major release.

The JDEM/Omega on-board software systems leverage from previous GSFC flight software.

With respect to the spacecraft flight software, the architecture follows directly from the product line of FSW systems created and flown at GSFC in the past decade. The GSFC Core Flight Software Executive (cFE) provides the fundamental framework which currently exists on missions such as LRO, SDO, and GPM, and will again be the backbone software system running on the RAD750/VxWorks environment. GSFC intentionally builds their spacecraft FSW systems to be readily tailored to meet the custom mission-unique requirements. By utilizing an existing robust and flexible software communications bus, functional building blocks may be readily included or removed to meet mission unique requirements. This customizable framework, which readily permits use of proven flight software functions both, reduces risk and cost by shortening the overall development cycle.

To support the high volume data transfers from the Science Data Recorder to the ground, the GSFC software/hardware team is considering using the established CFDP (CCSDS File Delivery Protocol) developed for previous and existing missions such as LRO, JWST and MMS. The same CFDP engine is used both on the ground system and the flight side and provides simple, robust, reliable data transfers to and from the spacecraft.

Fine Guidance Sensor (FGS) processing is distributed between the FGS focal plane electronics and the main processor in the spacecraft. After settling at a new target and achieving star tracker-level attitude performance, the Bus FSW commands the FGS to acquire stars. The FGS processing then isolates on observed stars in the FOV and autonomously goes into a fast read-out mode on a small set of pixels surrounding and including the stars. It performs fast read-out of the set of pixels associated with all observed stars and computes star centroids and intensities for each observed star followed by star identification using a star catalogue. The FGS then performs compensations for physical phenomena such as velocity aberration, parallax (if needed), proper motion, and detector-specific calibrations before computing the optimal. The FGS will also output quaternion statistical and quality information that the Bus FSW will utilize in weighting individual measured attitude quaternions in downstream processing.

On the instrument side, a RAD750 / VxWorks solution is also planned as the platform on which to host the flight software. The Instrument FSW (IFSW) will leverage off the cFE architecture (similar to the spacecraft FSW), as implemented on the JWST Instrument controller (ISIM). The IFSW plans to implement custom interfaces to the mission-unique hardware suite.

CMMI-certified software life cycle processes for spacecraft and instrument flight software are the cornerstone for the reliable and successful FSW development and test. Using well-documented industry-standard configuration management and software quality assurance processes and tools allows the FSW development team to move forward quickly and efficiently with respect to cost and schedule while keeping the managed risks at very low levels.

The development and test teams are provided with their own desktop level (one per developer) simulators as well as a high-fidelity simulator built around engi-neering model hardware (called a Flatsat) to ensure successful testing of the targeted software loads. The general process of software delivery consists of a series of 'Builds', each providing additional capabilities to the I&T team as they build up the spacecraft and/or instrument hardware during the integration phase. The final Build serves as the software system which is formally verified by test on the Flatsat to ensure that all requirements are met.

The Flatsat comprises a fully redundant (flight-like) set of engineering model hardware for the electronics and a dynamic simulator to provide a high fidelity simulation of the sensors and actuators. Flight software loads, as well as flight hardware components are formally checked out in the Flatsat environment prior to installation on the spacecraft.

There is no JDEM/Omega specific ground station development required. The DSN is currently in the process of updating its standard services to include Ka-band at 150 Mbps and this institutional upgrade will be in place and operational well before JDEM/Omega launches.

Science development will concentrate on the definition and implementation of the algorithms that are needed to properly calibrate the data for use by the science teams and the general scientific community and to perform the detailed scientific analysis. This will be an iterative process between the JDEM Science Teams and the SOC and includes the definition and implementation into actual software of detailed calibration requirements on basic items such as photometric accuracy, flat-fielding, and mosaicing of images and of higher level science data processing, such as identification and isolation of supernova events, extraction and calibration of their spectra and their time evolution, fits to galaxy shapes for Weak Lensing experiments, and very careful reduction of the NIR data for the BAO studies. See the response to Mission Operations #4 for further description of the science development.

3. Provide entries in the mission design table. For mass and power, provide contingency if it has been allocated. If not, use 30% contingency. To calculate margin, take the difference between the maximum possible value (e.g. launch vehicle capability) and the maximum expected value (CBE plus contingency).
See *Table 8*.

4. Provide diagrams or drawings showing the observatory (payload and s/c) with the instruments and other components labeled and a descriptive caption. Provide a diagram of the observatory in the launch vehicle fairing indicating clearance.
See *Figure 5*, *Figure 6*, *Figure 7* and *Figure 8*.

5. For the mission, what are the three primary risks?
Thus far, two mission risks (risks numbered 1 and 2) are being carried in *Table 13* in the Programmatic & Schedule section. The rest of the risks are being carried under the payload instrumentation, reflecting that this is where most of the mission unique developmental activity is taking place.

*Figure 5 – Observatory Configuration*

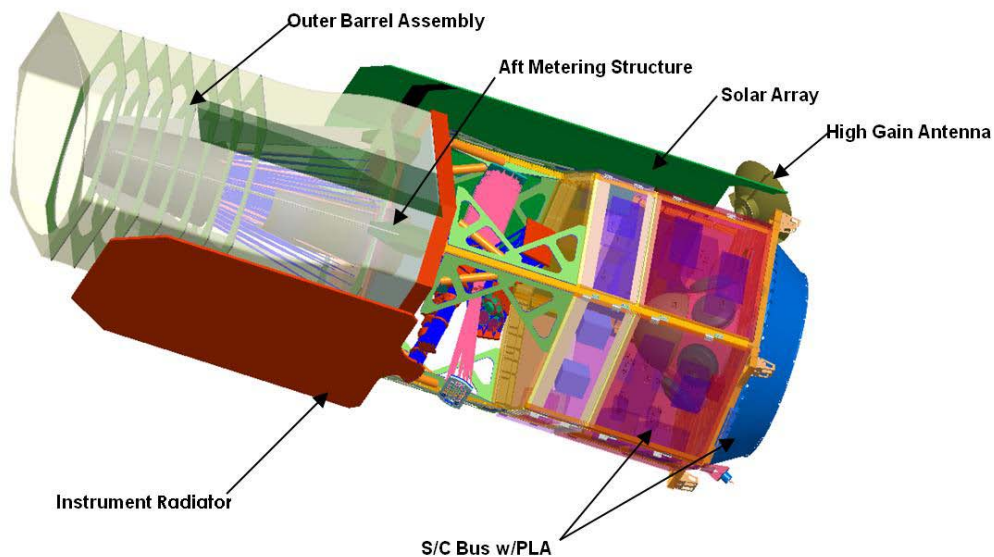

Figure 6 – JDEM/Omega Instrument Optical Layout

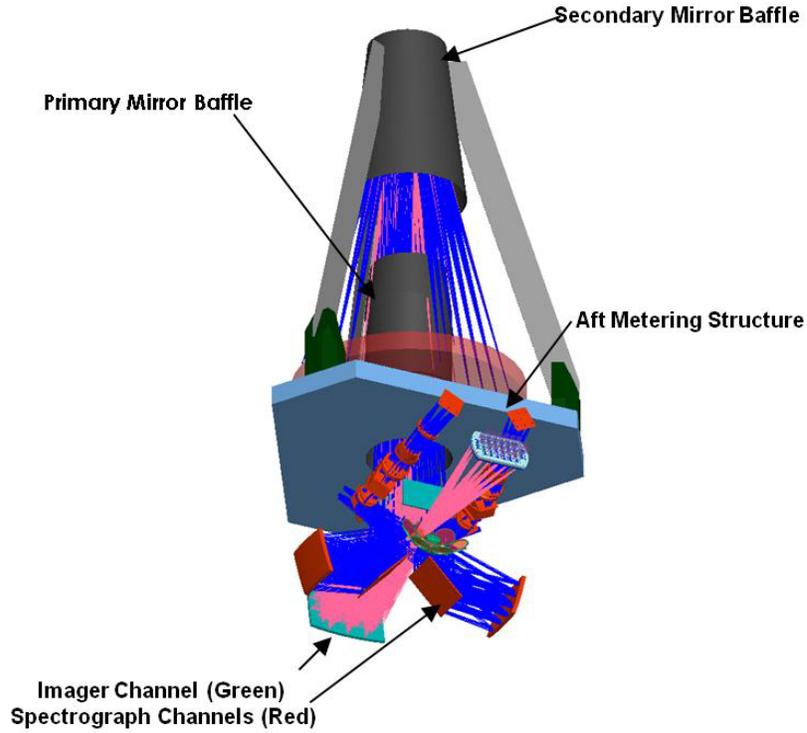

Figure 7 – JDEM/Omega Instrument Optical Channels Separated for Clarity

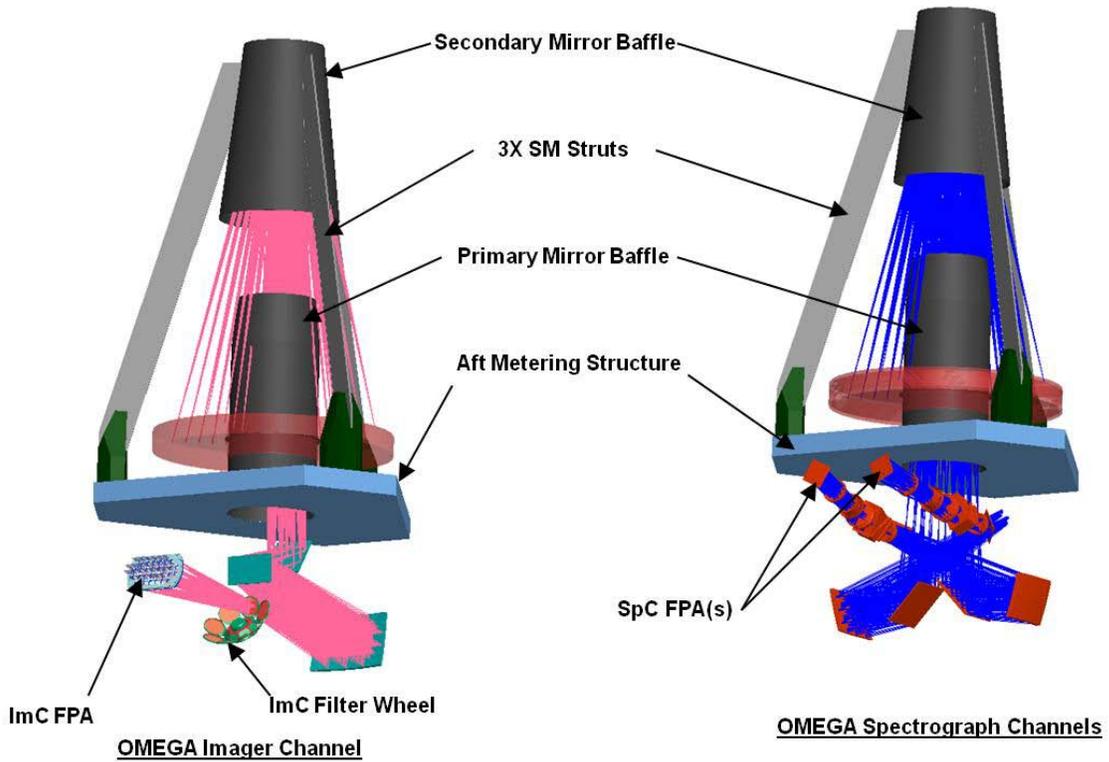

*Figure 8 – JDEM/Omega in the Launch Vehicle Fairing*

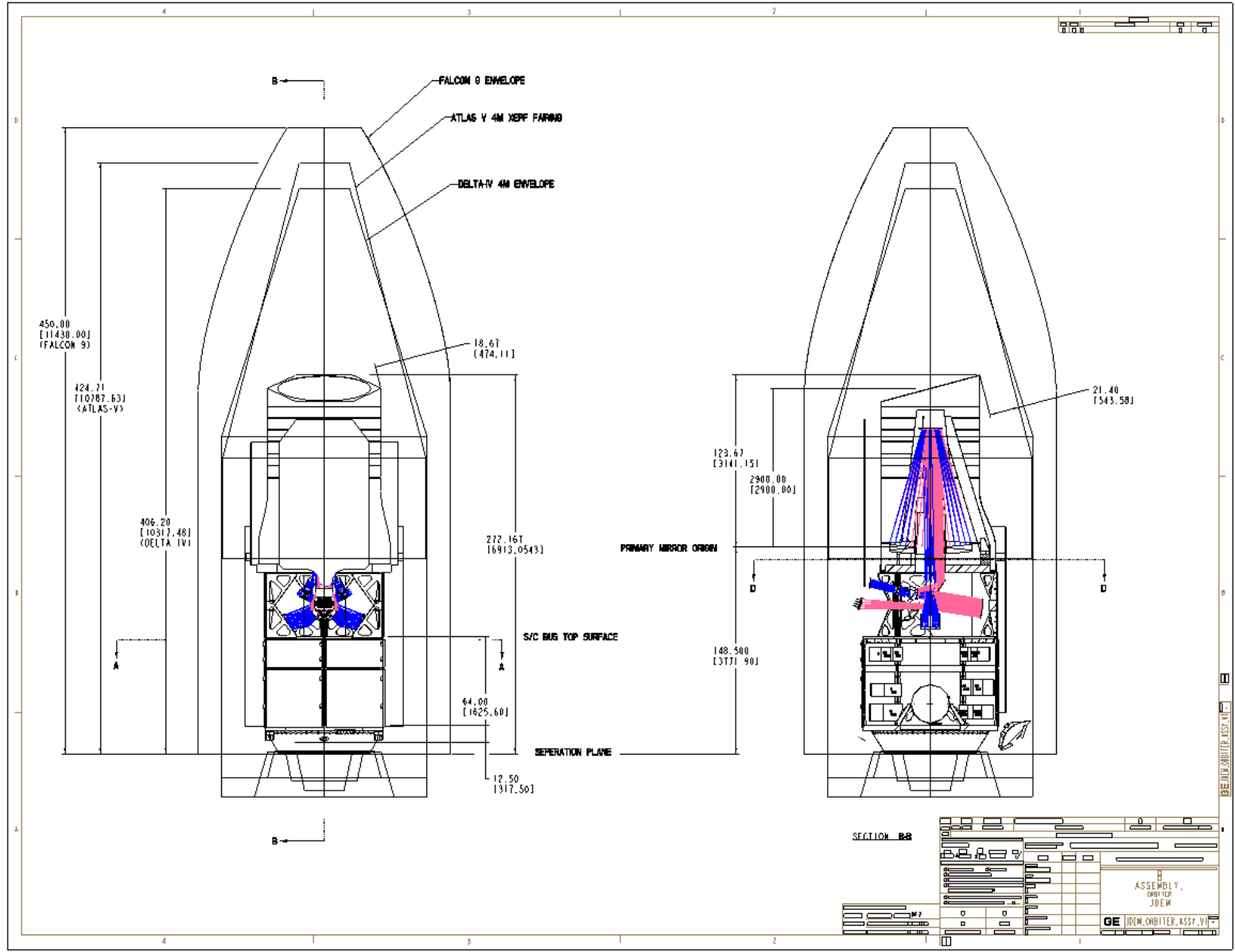

*Table 8 - Mission Design Table*

| Parameter | Value | Units |
|---|---|---|
| Orbit Parameters (apogee, perigee, inclination, etc.) | 10˚ x 28˚ Earth-Sun L2 Libration point orbit | |
| Mission Lifetime | 5 | yrs |
| Maximum Eclipse Period | None at L2 | min |
| Launch Site | CCAFS | |
| Observatory Dry Bus Mass **without** contingency | 1871 | kg |
| Observatory Dry Bus Mass contingency | 30 | % |
| Observatory Dry Bus Mass **with** contingency | 2424 | kg |
| Observatory Propellant Mass **without** contingency | 144 | kg |
| Observatory Propellant contingency | 30 | % |
| Observatory Propellant Mass **with** contingency | 187 | kg |
| Launch Vehicle | EELV (Atlas V, Delta IV) | Type |
| Launch Vehicle Mass Margin (Lift Capability - PAF - Observatory with Contingency) | Delta IV capability = 3123 Margin = 512 kg Atlas has higher capability | kg |
| Launch Vehicle Mass Margin (%) (Launch Vehicle Mass Margin/Observatory with Cont.) | Delta IV: 20% | % |
| Observatory Power **without** contingency | 1413 | W |
| Observatory Power contingency | 30 | % |
| Observatory Power **with** contingency | 1837 | W |

## Spacecraft Implementation

1. Describe the spacecraft characteristics and requirements. Include a preliminary description of the spacecraft design and a summary of the estimated performance of the key spacecraft subsystems. Please fill out the Spacecraft Mass Table.

The spacecraft design for JDEM/Omega is based on the Solar Dynamics Observatory (SDO) spacecraft, which was designed, manufactured, tested and qualified at GSFC. The spacecraft bus design provides cross strapping and/or redundancy for a single-fault tolerant design. *Structures:* The spacecraft bus is an aluminum hexagonal structure, consisting of two modules (bus module and propulsion module) which house the spacecraft & payload electronics boxes and the propulsion tank. The spacecraft bus provides the interfaces to the payload and the launch vehicle. It supports a 3-panel fixed solar array. *Attitude Control:* The spacecraft is three-axis stabilized, inertial and uses data from the fine guidance sensor, inertial reference unit and star trackers to meet the coarse pointing control of 3 arcsec RMS per axis, fine relative pointing control of 25 mas RMS per axis pitch/yaw, 1 arcsec roll and knowledge of 4 mas pitch/yaw and 300 mas roll. There are 2 fine guidance sensors with one located on the primary axis of the telescope and a pair of redundant sensors near the imager focal plane. The location of these sensors has been specifically chosen to optimize pointing performance through placement near the optical path, while also considering the influences of the thermal/mechanical environment. A set of 4 pyramidal reaction wheels is used for slewing as well as momentum storage. The attitude control electronics (ACE) box provides an independent firmware safe hold capability (using coarse sun sensors), to keep the observatory thermally-safe, power-positive and to protect the optical instruments from direct sunlight. *Propulsion:* A hydrazine mono-prop subsystem is required for orbit insertion, orbit maintenance and momentum dumping from the reaction wheels throughout the duration of the mission. *Electrical Power:* The power subsystem utilizes three fixed, body-mounted solar array panels to provide power for a daily average of ~1400 watts power using an 80 A-hr battery and a power supply electronics box. The solar array is currently sized to provide full Observatory power at EOL with 2 strings failed at the worst case observing angles. *Communications:* The communications subsystem uses S-band transponders to receive ground commands and to send real-time housekeeping telemetry to the ground via 2 Omni antennas as well as for ranging. A Ka-band transmitter with a gimbaled antenna will downlink stored science and housekeeping data at a rate of 150 Mbps without interrupting science operations. *Command & Data Handling:* The command and data handling subsystem includes a 1.1 Tb solid state recorder (SSR) sized to prevent data loss from a missed contact. The daily data volume is estimated at 0.8 Tb per day, assuming 2:1 lossless compression. The CDH/FSW provides fault management for the spacecraft health and safety as well as being able to safe the payload when necessary. *Thermal:* The spacecraft thermal design is a passive system, using tape, surface coatings, heaters and radiators. See *Table 10* for the Spacecraft Mass table. *Figure 9* shows the layout of the spacecraft subsystems inside the bus.

| Mission Lifetime | 5 years |
|---|---|
| Reliability | R = 0.85<br>Single Fault Tolerant |
| Data Rate | 150 Mbps stored data<br>1.1 Tb recorder |
| Pointing | Control: 25 mas P/Y, 1 a-s RMS<br>Knowledge: 4 mas P/Y 300 mas RMS |
| Power | 2500 W BOL solar array capability |
| Prop | Orbit adjusts to L2, maintain L2 orbit,<br>momentum unloading, EOM orbit |

2. Provide a brief description and an overall assessment of the technical maturity of the spacecraft subsystems and critical components. Provide TRL levels of key units. In particular, identify any required new technologies or developments or open implementation issues.

*Table 9* provides a description of the key spacecraft subsystem components, their heritage and their current technical maturity. All of the spacecraft components were selected based on their flight-proven and successful application on other missions.

3. Identify and describe the three lowest TRL units, state the TRL level and explain how and when these units will reach TRL 6.

Most spacecraft components are TRL 9, and all are at least TRL 6.

4. What are the three greatest risks with the S/C?

Due to the robustness and maturity of the spacecraft design and components, the Project is not carrying any risks on the spacecraft. As the program progresses, the spacecraft will be monitored for any risks that may arise.

5. If you have required new S/C technologies, developments or open issues describe the plans to address them (to answer you may provide technology implementation plan reports or concept study reports).

The JDEM/Omega spacecraft requires no new technologies.

6. Describe subsystem characteristics and requirements to the extent possible. Describe in more detail those subsystems that are less mature or have driving requirements for mission success. Such characteristics include: mass, volume, and power; pointing knowledge and accuracy; data rates; and a summary of margins. Comment on how these mass and power numbers relate to existing technology and what light weighting or power reduction is required to achieve your goals.

The JDEM/Omega spacecraft has been designed to provide all the resources necessary to support a telescope at L2 using mature and proven technology. Mass and power performance requirements are consistent with the current technology to build the spacecraft as well as launch on an EELV with more than adequate mass margin of over 30%. The solar array and its structure fit within the spacecraft structure needed to support their mass and meet MOI constraints. Gallium Arsenide solar array cells provide 28% efficiency and provide 2500 watts of output for an average orbit usage of ~1400 W. The remainder of the power subsystem is comprised of redundant power distribution units that control the distribution of power and provide unregulated 28 Vdc power to the payload.

The GNC system is comprised of off the shelf equipment for the gyro, star trackers, reaction wheels and coarse sun sensors. The fine guidance sensors will need to be customized for JDEM/Omega. The fine guidance sensor (FGS) is used to meet the fine pointing requirements needed for the WL and SN techniques. The primary Outrigger FGS consists of two pairs of HgCdTe detectors, a prime and redundant, located on outriggers on the imager FPA and fed through the imager optical train, including the filter wheel. This guider is used in all observations that include imaging. An additional pair of on-axis detectors, the On-Axis FGS, is fed from a separate field at the telescope intermediate focus via a separate optical train, which is unfiltered, and is used during imager spectroscopy. The On-Axis guider, along with the star trackers, provides pointing control for science observations when the imager is performing spectroscopy.

The prop subsystem does not have any unique features for JDEM/Omega. This subsystem is well within the requirements of other propulsion systems that have launched.

The C&DH subsystem will use the RAD750/cPCI architecture for controlling the observatory and interfacing with other components. The solid state recorder volume of 1.1 Tb has been used extensively on other missions. The ACE is a separate hardware box that is used in the event that the C&DH primary processor and algorithms cannot control the spacecraft. The ACE relies on coarse sun sensors to find the Sun and maintain a safe attitude with the instruments off.

The communications system is comprised of off the shelf hardware (e.g. Ka-band transmitter, S-band transmitter) and has no requirements that drive the hardware. The planned hardware is compatible with the DSN and SN.

7. Describe the flight heritage of the spacecraft and its subsystems. Indicate items that are to be developed, as well as any existing hardware or design/flight heritage. Discuss the steps needed for space qualification.

*Table 9* provides a description of the key spacecraft subsystem components, their heritage and their current technical maturity. There are no developmental items in the spacecraft design. The spacecraft components are flight qualified and the spacecraft bus structure will be fully flight qualified over all environments.

8. Address to the extent possible the accommodation of the science instruments by the spacecraft. In particular, identify any challenging or non-standard requirements (i.e. Jitter/momentum considerations, thermal environment/temperature limits etc).

The unique challenge on the JDEM/Omega spacecraft is to provide the environment required for the WL shape measurement. The science to engineering requirements decomposition process is being used to fully define the spacecraft to payload interface requirements. The spacecraft uses a Fine Guidance Sensor to meet the payload pointing requirements. To ensure that the entire observatory controls the effects of jitter and thermal mechanical distortions, integrated modeling techniques are being employed throughout the mission design phase in order to properly analyze the optical, thermal/mechanical and pointing performance.

9. Provide a schedule for the spacecraft, indicate the organization responsible and describe briefly past experience with similar spacecraft buses.

The integrated mission schedule, containing the spacecraft schedule, is included in *Figure 15* in the Programmatics and Schedule section.

The Goddard Space Flight Center is responsible for providing the JDEM/Omega spacecraft and has 50 years of experience in the design, analysis, fabrication, integration, and testing of spacecraft buses. This experience spans the development of spacecraft buses in-house at Goddard and managing the development of spacecraft buses by industry as part of end-to-end space flight Projects led by Goddard. Just a few examples of space science missions utilizing spacecraft buses developed under Goddard's leadership include Fermi, Swift, WMAP, XTE, CGRO and COBE. The most recent in-house spacecraft buses developed by God-dard include LRO, which launched successfully in June 2009 and the SDO mission, which is scheduled for launch later this year. The proposed spacecraft bus design for JDEM/Omega is based on the SDO spacecraft bus design, which has successfully completed all integration and test activities at Goddard and has been delivered to Cape Canaveral for launch. Goddard's qualifications and record of success in the development of spacecraft buses for scientific missions are unsurpassed.

10. Describe any instrumentation or spacecraft hardware that requires non US participation for mission success.

There are no instrumentation or spacecraft hardware components that require non US participation for mission success.

11. Fill out the Spacecraft Characteristics Table.
See *Table 11.*

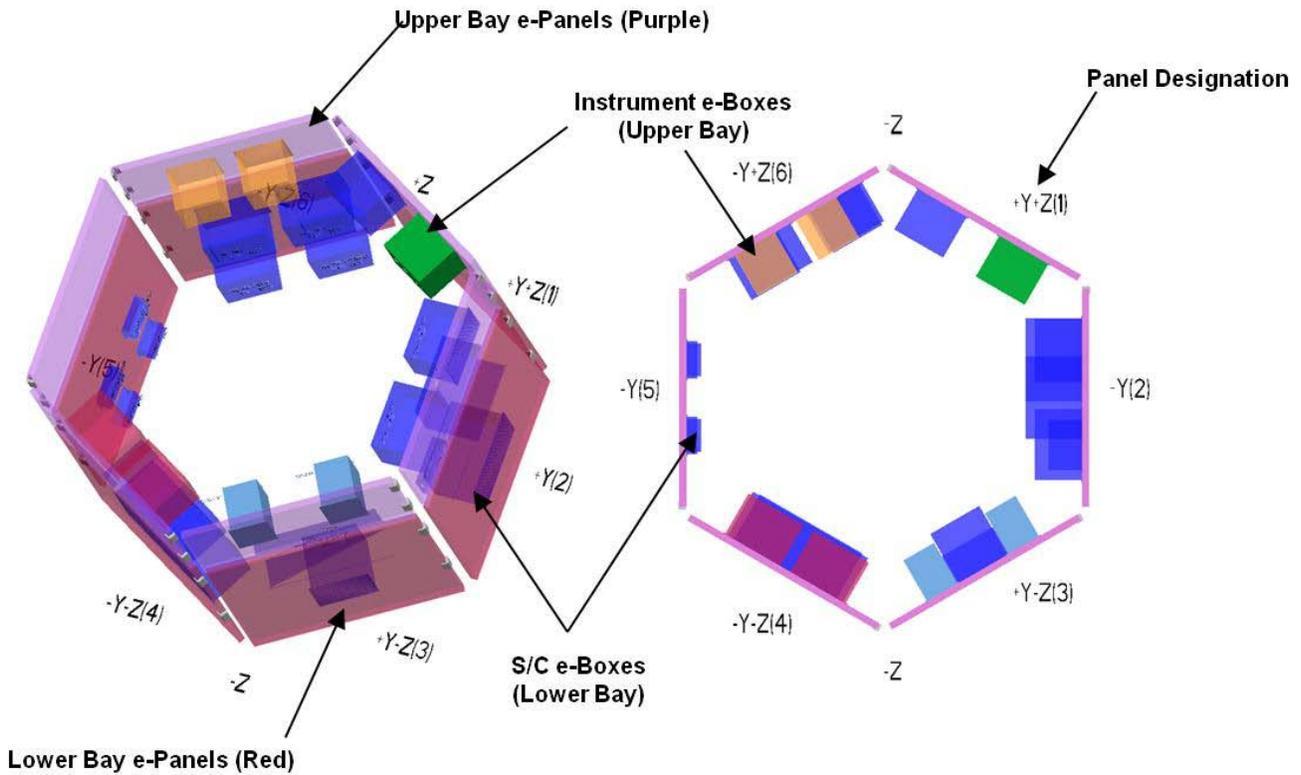

Figure 9 – Spacecraft Bus Configuration

*Table 9 - Heritage and Technology Readiness of JDEM/Omega Spacecraft Subsystem Key Components*

| Equipment | Features | Heritage | TRL |
|---|---|---|---|
| **C&DH Components** | | | |
| C&DH System Circuit Cards | Processor, cPCI backplane, MIL-STD-1553B. Includes cards that deliver 150 Mbps data rate from the SSR to the Ka band transmitter | LRO, SDO | TRL 6 |
| Solid State Recorder | 1.1 Tb capacity. | Worldview | TRL 9 |
| **COMM Components** | | | |
| Ka Band Modulator & Exciter | 150Mbps 26.5 GHz | LRO, SDO | TRL 9 |
| Ka Band 30W TWTA | Design is capable of 20 to 200W power output. JDEM/Omega device will be tuned to operate at 30W | SDO, LRO | TRL 9 |
| S Band Transponder | Used for cmd & non-science tlm downlinks (45W) | SDO | TRL 9 |
| S Band 20W TWTA | Used for cmd & non-science tlm downlinks | SDO | TRL 9 |
| **EPS Components** | | | |
| Battery | 80 A-hr (Li-ion), double deck modular construction | LRO, SDO | TRL 9 |
| Solar Cells | 3 panel (~14 m²), TJGaAs cells, honeycomb panel – composite facesheet, Al core w/ Kapton layer | SDO | TRL 9 |
| Power Supply Electronics | 28 vDC | LRO, SDO | TRL 6 |
| **ACS Components** | | | |
| Star Trackers | Solid-state, capability to track multiple stars | Terra, Fermi | TRL 9 |
| Inertial Reference Unit | High-precision | NEAR, Fermi | TRL 9 |
| Reaction wheels | High momentum storage, high torque, low disturbance | TRMM, XTE, WMAP | TRL 9 |
| Sun Sensors | Coarse or fine are acceptable | Swift, Rhessi, SDO | TRL 9 |
| Attitude Control Electronics | Independent safemode firmware computer | SDO, LRO | TRL 6 |
| HGA gimbal | Two-axis | SDO | TRL 9 |
| FGS | Two focal plane assemblies used to track multiple guide stars for precision pointing | JWST | TRL 6 |
| **PROP Components** | | | |
| Tank | Titanium | ERBS | TRL 9 |
| Thrusters | (8) 5 lbf – monoprop (N2H4) | LRO | TRL 9 |
| **Thermal Components** | | | |
| Heaters | Kapton film heaters | Numerous missions | TRL 9 |
| Temp Sensors | Thermistors, RTDs | Numerous missions | TRL 9 |
| Thermostats | Bi-metalic thermostats | Numerous missions | TRL 9 |

*Table 10 - Spacecraft Mass Table (kg)*

| Spacecraft bus | Current Best Estimate (CBE) | Percent Mass Contingency | CBE Plus Contingency (kg) |
|---|---|---|---|
| Structures & Mechanisms | 388 | 30 | 504 |
| Thermal Control | 42 | 30 | 55 |
| Propulsion (Dry Mass) | 38 | 30 | 49 |
| Attitude Control | 124 | 30 | 161 |
| Command & Data Handling | 52 | 30 | 68 |
| Telecommunications | 44 | 30 | 57 |
| Power | 170 | 30 | 221 |
| Total Spacecraft Dry Bus Mass | 858 | 30 | 1115 |

Table 11 - Spacecraft Characteristics Table

| Spacecraft bus | Value/ Summary, units |
|---|---|
| **Structure** | |
| Structures material (aluminum, exotic, composite, etc.) | Aluminum |
| Number of articulated structures | 1 – Ka-Band Antenna |
| Number of deployed structures | 1 – Ka-Band Antenna |
| **Thermal Control** | |
| Type of thermal control used | Passive – thermal coatings, MLI, heaters |
| **Propulsion** | |
| Estimated delta-V budget, m/s | 120 |
| Propulsion type(s) and associated propellant(s)/oxidizer(s) | blowdown, monoprop (hydrazine) system |
| Number of thrusters and tanks | 8 thrusters, 1 tank |
| Specific impulse of each propulsion mode, seconds | 220 |
| **Attitude Control** | |
| Control method (3-axis, spinner, grav-gradient, etc.). | 3-axis |
| Control reference (solar, inertial, Earth-nadir, Earth-limb, etc.) | Inertial |
| Attitude control capability, milli-arcseconds | 25 mas (pitch/yaw), 1 (roll) a-s |
| Attitude knowledge limit, milli-arcseconds | 4 (pitch/yaw), 300 (roll)  mas |
| Agility requirements (maneuvers, scanning, etc.) | 0.7 deg in 38 secs (includes both slew and settle) |
| Articulation/#–axes (solar arrays, antennas, gimbals, etc.) | dual axis, Ka-band gimbal |
| Sensor and actuator information (precision/errors, torque, momentum storage capabilities, etc.) | STA (2 a-s accuracy, track up to 8 deg/sec) RWA (50 Nms, 0.3 Nm) SIRU (AWN: 1 mas/root-Hz, ARW: 36 mas/root-hr) FGS (0.1 deq square FOV detectors; centroiding 16th magnitude guide stars to 4 mas accuracy) |
| **Command & Data Handling** | |
| Spacecraft housekeeping data rate, kbps | 8 kbps |
| Data storage capacity, Tbits | 1.1 Tb |
| Maximum storage record rate, Mbps | 300 Mbps |
| Maximum storage playback rate, Mbps | 300 Mbps |
| **Power** | |
| Type of array structure (rigid, flexible, body mounted, deployed, articulated) | Rigid, body mounted panels |
| Array size, meters x meters | 14 |
| Solar cell type (Si, GaAs, Multi-junction GaAs, concentrators) | Triple Junction GaAs |
| Expected  power generation at Beginning of Life (BOL) and End of Life (EOL), watts | >2500 (BOL) >2050 (EOL) |
| On-orbit average power consumption, watts | ~1400 |
| Battery type (NiCd, NiH, Li-ion) | Li-ion |
| Battery storage capacity, amp-hours | 80 |

## 3. ENABLING TECHNOLOGY

*Please update or provide information from the original RFI response describing new Enabling Technologies that must be developed for mission success.*

JDEM/Omega benefits from significant development and testing activities on other GSFC programs, notably the Wide Field Camera 3 (WFC3) instrument for the Hubble Space Telescope (HST) and JWST, for the critical technology used. Two key technologies are used on JDEM/Omega, but due to the previous work done, *no new technology development is needed.* Additionally, no non-U.S. technology is required. Ongoing Project efforts geared toward the application of these proven technologies to the JDEM/Omega instrument are described below.

### Near-Infrared Detectors

The HgCdTe near-infrared detectors baselined have extensive and direct heritage from JWST. These detectors are exact copies of the JWST short-wave detectors, and are the result of over 5 years of development. The dark current performance requirements for JDEM/Omega are less stringent than those for JWST by at least one order of magnitude because of the relatively high Zodiacal background for slitless spectroscopy and broadband imaging. JWST has already demonstrated IR detector performance to the levels required for JDEM, and JDEM will not set specifications beyond what has already been demonstrated.

The number of detectors required for JDEM/Omega is large compared to previous flight programs (but not ground arrays), but the manufacturing capabilities for these detectors have improved sufficiently that this does not pose a significant risk. Over the course of the JWST program, Sensor Chip Assembly (SCA) yields have substantially improved, and this is now demonstrated routinely with the similar, science-grade SCAs manufactured for other programs. The assumed yields used for costing the JDEM/Omega program incorporate these most recent yield data.

JDEM/Omega uses a 6x4 mosaic of these SCAs. Current experience with JWST includes up to 2x2 mosaics, see *Figure 10.* Implementing the 6x4 mosaic required for JDEM/Omega is possible using either an extension of the JWST design, or by drawing from the technologies used in constructing the very large ground and space-based focal planes currently in progress. The technologies used for these mosaic architectures are at TRL6. A JDEM/Omega Engineering Develop-ment Unit (EDU) FPA is being started early to minimize the risk of scaling the FPA size.

The direct use of the JWST 2.5 µm cutoff detectors at temperatures below ~75 K is at TRL7. Initial results from JDEM detector characterization testing indicates these detectors can be operated up to 100 K, allowing a reduced FPA radiator.

Another possibility is being explored that may simplify the thermal and optical system design for the instruments. The scientific requirements do not extend longwards of 2.0 µm, yet the baseline detectors have sensitivity up to 2.5 µm. This additional bandpass increases complexity because it causes additional dark current at any given temperature and also additional background from in-band thermal radiation (from the optics). The former is mitigated by larger radiators and the latter with shortpass filters. Instead of creating a 2.0 µm cutoff instrument using a 2.5 µm cutoff detector, it is possible to construct the detector so that its intrinsic cutoff is ~2.0 µm. The development risk is small because of the extensive development that has already taken place for the WFC3 instrument infrared channel and by the early development work for the Supernova Acceleration Probe (SNAP), both of which are 1.7 µm cutoff. Interpolating between 1.7 µm and 2.5 µm is a relatively low-risk modification to existing processes, should it be deemed desirable by upcoming mission-level trades.

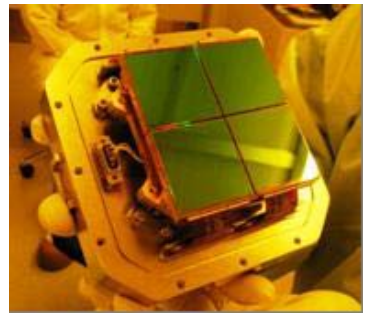

*Figure 10 - JWST NIRCam-style 2x2 mosaic of 2Kx2K 2.5 µm cutoff HgCdTe SCAs.*

### Cryogenic Lenses

The large refractive lenses used at cryogenic temperatures (~140K) in the NIR spectrometer have heritage at GSFC from the Composite Infrared Spectrometer (CIRS) on the Cassini mission and on the Infrared Array Camera (IRAC) on Spitzer. JDEM/Omega has begun activities to mitigate the risk associated with scaling to the larger JDEM/Omega cryogenic lens mounts.

The design includes only previously flown, radiation hard refractive materials (e.g. fused silica, $CaF_2$ and ZnSe). The narrower bandpass for JDEM allows for the selection of these more robust lens materials than for broader bandpass instruments such as JWST NIRCam. These materials have strong flight heritage and GSFC has extended precise refractive index knowledge of all of these materials to cryogenic temperatures for JWST. JDEM/Omega will benefit from the cryogenic optical systems expertise built up from Spitzer/IRAC, Cassini/CIRS, and JWST.

Risk reduction activities are planned for each perceived risk area. Potential risk areas for cryogenic lenses at L2 include fabrication risks (e.g. aspheric polishing, coating stress), mounting stress, launch survival, cryogenic wavefront performance (including both mounting and temperature effects to 140K) and system alignment.

The first activity will be to polish a set of windows of each material, mount them in JDEM/Omega designed mounts and perform environmental testing (vibration, acoustics and cryogenic cycle testing) with wavefront remeasurments at each step. In parallel, GSFC will polish and coat pathfinder lenses and prisms to demonstrate the ability to fabricate to the required ambient wavefront quality.

These early project activities will be followed by a full engineering test unit of the NIR spectrometer channel to prove out each component as well as instrument alignment. As the NIR spectrometer works from a collimated beam, these ETUs are readily testable without a full telescope simulator.

1. For any technologies rated at a Technology Readiness Level (TRL) of 5 or less, please describe the rationale for the TRL rating, including the description of analysis or hardware development activities to date, and its associated technology maturation plan.

No technologies are rated below TRL 6.

2. Describe the critical aspect of the enabling technology to mission success and the sensitivity of mission performance if the technology is not realized.

A part of the power of the JDEM/Omega configuration comes from using the NIR SCA's in its imaging channel to perform Weak Lensing measurements. While we believe that this is very possible, there is very little experience with using these detectors for this purpose in existing ground-based or space-based measurements.

We have started a program to examine the effects of known imperfections in the HgCdTe detectors on shape measurements. A robust set of simulation activities based on, and in parallel with, the work on CCD's will link the detector performance levels to shape sensitivity. Laboratory measurements will feed this simulation mechanism with representative data and their stability to provide the overall impact to the measured shapes. It is possible that the HgCdTe devices may turn out to be more desirable because the systematic errors are stable with time. For CCD's this is not the case, since the main feature that affects the shapes is the Charge Transfer Efficiency, which degrades with time on orbit. Areas to be studied include Inter-Pixel Capacitance, Intra-Pixel Response, Persistence, short wavelength QE and DQE, and reciprocity failure (where short exposures to bright light does not give the same answer as long exposures to dim light). These activities have started, and preliminary results are expected by the end of the year.

3. Provide specific cost and schedule assumptions by year for Pre-Phase A and Phase A efforts that allow the technology to be ready when required.

As described above, early activities in Pre-Phase A and Phase A are for early risk reduction work as opposed to work to bring new technology up to TRL 6; therefore, no new technology funds are required.

# 4. MISSION OPERATIONS DEVELOPMENT

1. Provide a brief description of mission operations, aimed at communicating the overall complexity of the ground operations (frequency of contacts, reorientations, complexity of mission planning, etc). Analogies with currently operating or recent missions are helpful. If the NASA DSN network will be used provide time required per week as well as the number of weeks (timeline) required for the mission.

The ground system is comprised of 4 major elements: the Mission Operations Center (MOC); the Science Operations Center (SOC); the Science Support Center (SSC) and a Data Archive. The SSC is responsible for science planning and scheduling. It receives inputs from the SOC for special activities as well as routine activities. It produces a 7 day science schedule that is sent to the MOC, where the weekly mission plan, schedule, and command loads are generated. Activities such as DSN scheduling and housekeeping activities are added by the MOC. Developing the schedule is an iterative process. Because JDEM/Omega is primarily a survey mission, both long term and short term weekly plans are completed well in advance of execution. The planning and scheduling system takes the long-range plans for the observing cadences and fits them into one-week planning windows. The detailed schedule for the observations, the management of on-board data storage and its transmission to the ground (done automatically) is produced along with a weekly command load for execution. Command loads are built weekly and expected to be uplinked once a week. Minor activities (e.g. dumping computer diagnostic data) requiring commanding will be done during week days. JDEM/Omega planning is very similar to Fermi: weekly uplinks, observing repeatability, an occasional deviation and a "mostly by exception only" scheduling changes.

The MOC, SOC, SSC and data archive are staffed on an 8-hour, 5 days a week basis. Health and safety monitoring of the observatory is done autonomously by the MOC by examining the real time and recorded data for limit violations or unexpected configurations. Depending upon the severity level of the limit violation, the MOC sends an alert to appropriate personnel for follow on action. The SOC also monitors instrument housekeeping data. Since the observatory is required to monitor and safe itself and the payload in the event of an anomaly, the ground needs only to monitor observa-tory telemetry, but not send any commands autonomously.

Data processing is done automatically by the ground system. Once the recorded science and housekeeping data is received at JPL from the DSN, it is transmitted to the MOC where it is level 0 processed and stored. The MOC then sorts the data and distributes it automatically to the SOC and the data archive. Each SOC receives only the data it needs. Once the SOC receives the data, higher level processing (levels 1-3+) begins. This is done automatically with manual intervention required only when data is missing or corrupted. Scientists and operations specialists staff the SOC to resolve these problems, assist users as well as evaluate instrument performance, perform calibrations and maintain instrument flight software.

Approximately 400 Gbits of recorded science and housekeeping data are automatically transmitted to the ground twice a day (for a total of ~800 Gbits) using the DSN stations, with each contact lasting no more than 2 hours. The DSN is required throughout the lifetime of the mission. At least 30 minutes of contact time is reserved in the unlikely event there are problems somewhere in the down link chain. For example, if a DSN contact were cancelled after it had been scheduled, there is no need to schedule an additional contact to prevent data loss as there is adequate data storage on-board to accommodate missed contacts. The next few contacts would be sufficient to catch up.

2. Identify any unusual constraints or special communications, tracking, or near real-time ground support requirements.

There are no unusual constraints, special communications, tracking, or near real-time ground support requirements. S-band will be used for the uplink and downlink with Ka-band (150 Mbps) being used to playback stored science and housekeeping data. The DSN is currently in the process of updating its standard services to include Ka-band at 150 Mbps and these improvements will be in place well before JDEM/Omega launches. Scheduling any DSN assets twice daily will satisfy all commanding and playback requirements. The mission requires 2-way tracking in order to meet the orbit determination requirements. Being at L-2 requires that tracking be performed at both a southern latitude and northern latitude site.

3. Identify any unusual or especially challenging operational constraints (i.e. viewing or pointing requirements).

There are no challenging constraints for operations as JDEM/Omega is a survey mission, with both wide mapping and deep field-monitoring surveys. The gimbaled antenna for the Ka-band allows stored data to be downlinked during a contact. Very small antenna slews may be required during repointings, but these will be done in conjunction with the spacecraft slews and therefore it will not impact the science. There will be momentum unloading using thrusters every 3-4 days that will impact pointing control, but they are of short duration and accounted for in observing efficiency estimates. The field of regard is restricted to +80 to +120 degrees from the Sun with roll off the Sun restricted to ±45°. This is easily accommodated in operations.

The pointing requirements are handled by the onboard systems and do not directly impact the operations in terms of being driving requirements. The pointing knowledge and control requirements are tight but achievable, and are addressed in the spacecraft section. The planning and scheduling systems include estimates for slew/settle times, which are autonomously adjusted on-orbit to accommodate the actual times via simple flag logic. For the BAO/WL survey, there are multiple opportunities to observe specific targets throughout the year while the SN survey requires 5 day revisits to its field. Being at L2 means there is no SAA and neither Earth avoidance constraints nor Earth eclipses of the Sun.

4. Describe science and data products in sufficient detail that Phase E costs can be understood compared to the level of effort described in this section.

The raw science data products comprise image data with multiple filter bands, roll angles, and/or epochs per sky field. The slitless spectroscopy data contains multiple exposures and roll angles per field. The data are Level 0 processed at the MOC and then sent to the appropriate SSC to be flat-fielded, corrected for geometric distortions, photometrically calibrated to provide absolute fluxes, and, in the case of spectra, wavelength calibrated. This level of calibration is sufficient for general astronomical use. This database will enable the production of enormous catalogs containing multi-band NIR photometry (with multiple epochs for the supernova fields), and spectroscopic redshifts for H-alpha (and some other line) emitting galaxies.

For specific dark energy studies using supernovae, weak lensing, and baryon acoustic oscillations, higher-level data products are needed. For supernova studies, the processing requires source identification from difference images, calibrated light curve extraction from the multi-epoch data, supernova type identification from the low resolution spectroscopy data, and luminosity estimation from fits to the type Ia light curve models. For weak lensing, the processing requires source identification, shape estimation, and photometric redshift determination for each object. The shape and redshift measurements will both require rigorous analysis programs to determine the statistical and systematic error limits associated with each measurement. This will be the most complex part of the weak lensing processing. For baryon acoustic oscillations, the analysis requires source identification and redshift determination, which requires collating and registering the imaging and spectral data. The most complex part of that processing will be the error analysis of the redshift measurement, including the level of confusion from overlapping spectra and the interloper rates from non-H-alpha line emission.

Science development will concentrate on the definition and implementation of the algorithms that are needed to properly calibrate the data for use by the science teams and the general scientific community and to perform the detailed scientific analysis. This will be a shared process between the JDEM Science Teams and the SOC's. The SOCs will be responsible for software tools to handle dataflow management, archiving and distribution, and for data reduction and calibration for the general scientific community. The JDEM science teams will be responsible for developing data processing software optimized for their unique analysis requirements.

The data reduction software development may be divided into several levels of data products. Low-level reduction applies basic corrections and calibrations to the data, such as instrument background removal, flat-fielding, correction for optical distortions, determining the absolute astrometric solution for each exposure, etc. These routines are likely to be common to all data analysis techniques. Mid-level data reductions will perform tasks such as combining multiple exposures of a field to produce sub-sampled images, correcting for image persistence in the detectors remaining from pre-vious exposures, assembling individual images into a full-sky image with corresponding exposure map, and combining multiple spectral images. Many of the mid-level routines will be common for all users, but some may need to be optimized for particular purposes. Some high-level data reduction software will be general in nature, such as: extraction of sources from the images, their characterization, and creation of a source catalog; this includes software to correlate sources in the JDEM/Omega catalog with those in other catalogs. Other high-level software will be unique to each data analysis technique. The SN program will include software to detect transient events and produce the corresponding light curves, extract the corresponding spectra from both the SN at various intervals and of the host galaxy at times when the SN is not present, and match templates to the SN spectra for typing and redshift determination. The WL program will include software for measuring galaxy shapes and determining photometric redshifts. The BAO program will include spectral extraction software optimized for faint sources, wavelength calibration, and identification of emission lines.

Calibration software will derive the necessary instrument characteristics from both ground and in-flight data. Examples include optical distortion mapping, flatfield and flux calibration, image persistence, and characterization of the PSF. All calibrations will have to be tracked as a function of time, instrument configuration, and environment. Software for managing the dataflow and archiving data products will be developed. Experience gained at STScI, SDSS, and Pan-STARRS will be applicable to many aspects of the JDEM/Omega data reduction effort.

Data analysis software must be developed to deduce the properties of dark energy from the reduced data. The software algorithms needed for the SN program are in a relatively mature state, and the computation demands are lower than for the other methods: the total number of SNe to be analyzed is small, and they may be analyzed individually. The BAO program needs algorithms to derive the matter power spectrum of 100 million galaxies distributed over a large cosmic volume. Such analyses have already been performed on datasets that are smaller than will be provided by JDEM/Omega, but which were in many ways more challenging; scaling these tools to the JDEM/Omega dataset will be straightforward. The WL data analysis software will need to derive the matter distribution from the shear and redshift measurements of up to ~1 billion galaxies. WL analysis algorithms are presently under active development by the WL science community. The Project will work closely with this community to adapt these algorithms for the JDEM/Omega dataset.

The Science Operations Center will have software to perform the detailed planning of the surveys, track the progress of the surveys, and plan calibration observations. Software to Tools for assessing data quality will be used to provide early feedback on the survey strategy and allow for modifications if they are needed.

The Project will take advantage of the expertise available from current large astronomical data processing efforts, as well as the data reduction experience currently being gained by the JWST project for the same detectors to be used by JDEM/Omega.

5. Describe the science and operations center for the activity: will an existing center be expected to operate this activity?; how many distinct investigations will use the facility?; will there be a guest observer program?; will investigators be funded directly by the activity?

The JDEM/Omega ground system consists of a Mission Operations Center (MOC), a distributed Science Operations Center (SOC), a Science Support Center (SSC) and a Data Archive. For each element, existing facilities and infrastructure will be leveraged to provide the maximum possible cost savings and operational efficiencies.

The MOC performs spacecraft, telescope and instrument health & safety monitoring, real-time and stored command load generation, spacecraft subsystem trending & analysis, instrument and telescope calibrations, spacecraft anomaly resolution, safemode recovery, level 0 data processing and transmission to the SOC and the SSC. With inputs from the SSC, it performs Mission-level Planning and Scheduling.

The SSC is responsible for science planning & scheduling, supporting mission planning activities carried out by the MOC, running the Participating Scientist Program, providing Science Team and Participating Scientist support, and performing EPO activities for the public and the astronomical community.

The SOC is responsible for generating level 1-3+ data products. They ingest Level 0 data from the MOC and perform Level 1-3 data processing for the Science Teams and the Participating Scientists and transmit these calibrated data to the SSC and from there to the Data Archive. They also provide science & planning inputs to the SSC, generate instrument command loads, and support instrument anomaly resolution.

Approximately 6 dedicated Science Teams will be funded over a 5-year period to execute the primary dark energy science programs. In this period, the Participating Scientist Program (PSP) for ancillary science provides additional funding for reduction and analysis of the JDEM/Omega data by non-dark energy astronomers. Operations costs and grants for the PSP in the primary mission are fully included in the lifecycle costs. If JDEM/Omega operations are extended beyond the 5-year baseline, the PSP becomes a Guest Observer program. These costs were not included in the lifecycle cost estimate as it only covered the baseline mission. We expect a total of 50 PSP/Ancillary Science investigations to be supported each year during the primary mission and anticipate additional PSP/GO investigations during any extended mission.

## 6. Will the activity need and support a data archive?

The activity requires an archive facility to ingest and archive Level 0-3+ data along with any higher-level data products produced by the Science Teams and PSP, manage proprietary data periods, provide data search and access tools, and distribute data to the science teams and astronomical community. The archive will most likely be based at an existing multi-mission archive center.

*Table 12 - Mission Operations and Ground Data Systems Table*

| Downlink Information | Value |
|---|---|
| Number of Contacts per Day | 2 |
| Downlink Frequency Band, GHz | 26.5 |
| Telemetry Data Rate(s), Mbps | 150 (science and housekeeping) |
| S/C Transmitting Antenna Type(s) and Gain(s), DBi | 0.75 m Ka band, gimbaled, 43 dBi |
| Spacecraft transmitter peak power, watts. | 108 watts DC power (modulator + TWTA) |
| Downlink Receiving Antenna Gain, DBi | DSN 34m G/T = 54.3 dB / degrees - K |
| Transmitting Power Amplifier Output, watts | 30 |
| Uplink Information | Value |
| Number of Uplinks per Day | 1 |
| Uplink Frequency Band, GHz | 2.1064 |
| Telecommand Data Rate, kbps | 2 |
| S/C Receiving Antenna Type(s) and Gain(s), DBi | 2 S-band Omnis 5.0 dBi max gain |
| AAAC | Astronomy and Astrophysics Advisory Committee |
| AANM | Astronomy and Astrophysics in the New Millennium |
| ACE | Attitude Control Electronics |
| ACS | Attitude Control System |
| ACTS | Advance Communications Technology Satellite |
| ADC | Analog-to Digital Converters |
| ADEPT | Advanced Dark Energy Physics Telescope |
| AEHF | Advanced Extremely High Frequency |
| AIM | Aeronomy of Ice in the Mesosphere |
| AO | Announcement of Opportunity |
| ASIC | Application Specific Integrated Circuit |
| ASSIST | Automated Satellite Support and Integration System Test |
| ATP | Authority to Proceed |
| BAE | British Aerospace |
| BAO | Baryon Acoustic Oscillations |
| BEI | Baldwin Electronics Incorporated |
| BEPAC | Beyond Einstein Program Assessment Committee |
| BOSS | Baryon Oscillation Spectroscopic Survey on SPSS |
| C&DH | Command and Data Handling |
| CaF2 | Calcium Fluoride |
| CALIPSO | Cloud-Aerosol Lidar and Infrared Pathfinder Satellite Observation |
| CBE | Current Best Estimate |
| CCAFS | Cape Canaveral Air Force Station |
| CCD | Charged Coupled Device |
| cCPI | Compact Computer Peripheral Interconnect |
| CCSDS | Consultative Committee for Space Data Systems |
| CDR | Critical Design Review |
| CFDP | CCSDS File Delivery Protocol |
| cFE | Core Flight Software Executive |
| CFS | Core Flight System |
| CGRO | Compton Gamma Ray Observatory |
| CIRS | Composite Infrared Spectrometer |
| CMB | Cosmic Microwave Background |
| CMMI | Capability Model Maturity Index |
| CMOS | Complementary Metal Oxide Simi-conductor |
| COBE | Cosmic Background Explorer |
| COTS | Commercial Off the Shelf |
| CY | Calendar Year |
| DBi | Decibels Isotropic |
| DCL | Detector Characterization Lab |
| DECS | Dark Energy Cosmology Satellite |
| DESTINY | Dark Energy Space Telescope |
| DETF | Dark Energy Task Force |
| DOE | Department of Energy |
| DOF | Degree of Freedom |
| DQE | Detective Quantum Efficiency |

| DSN | Deep Space Network |
| DTAP | Detector Technology Advancement Program |
| DWG | Detector Working Group |
| E(B-V) | Extinction (B-V) |
| e-Boxes | Electronic-Boxes |
| EDU | Engineering Development Unit |
| EE | Encircled Energy |
| EELV | Evolved Expendable Launch Vehicle |
| EMI/EMC | Electromagnetic Interference/Electromagnetic Compatibility |
| EOL | End of Life |
| EOM | End-of-Mission |
| EoM-E | End of Mission – Extended |
| EoM-P | End of Mission – Primary |
| e-Panels | Electronic-Panels |
| EPO | Education and Public Outreach |
| ERBS | Earth Radiation Budget Satellite |
| ESA | European Space Agency |
| ETU | Engineering Test Unit |
| FDC | Fault Detection and Correction |
| FGS | Fine Guidance Sensor |
| FMEA | Failure Modes and Effects Analysis |
| FMOS | Fiber Multi-Object Spectrograph |
| FOM | Figure of Merit |
| FoM | Figure of Merit |
| FoMSWG | Figure of Merit Science Working Group |
| FOR | Field-of-Regard |
| FOV | Field-of-View |
| FPA | Focal Plane Array |
| FPE | Focal Plane Electronics |
| FPGA | Field Programmable Gate Array |
| FSW | Flight Software |
| FY | Physical |
| G/T | Ground Terminal |
| GHz | Gigahertz |
| GNC | Guidance Navigation and Control |
| GO | Guest Observer |
| GOTS | Government Off the Shelf |
| GPM | Global Precipitation Mission |
| GSFC | Goddard Space Flight Center |
| H/W | Hardware |
| HGA | High Gain Antenna |
| HgCdTe | Mercury Cadmium Telluride |
| HST | Hubble Space Telescope |
| I&T | Integration and Test |
| ICDH | Instrument Command and Data Handling |
| ICE | Instrument Control Electronics |
| IFSW | Instrument Flight Software |
| IM | Integrated Modeling |
| ImC | Imager Channel |
| IPC | Interpixel Capacitance |
| IR | Infrared |
| IRAC | Infrared Array Camera |
| IRMOS | Infrared Multi-object Spectrometer |
| ISIM | Integrated Science Instrument Module |
| ITOS | Integrated Test and Operation System |
| JAXA | Japan Aerospace Exploration Agency |
| JCL | Joint Confidence Level |
| JDEM | Joint Dark Energy Mission |
| JHU | Johns Hopkins University |
| JPL | Jet Propulsion Laboratory |
| JWST | James Webb Space Telescope |
| Kbps | Kilobits Per Second |
| KDP | Key Decision Point |
| KSC | Kennedy Space Center |
| LCCE | Lifecycle Cost Estimate |
| LDCM | Landsat Data Continuity Mission |
| LDR | Launch Readiness Date |
| Li-ion | Lithium-ion |
| LISA | Laser Interferometer Space Antenna |
| LRO | Lunar Reconnaissance Orbiter |
| LVDS | Low Voltage Differential |
| Mbps | Megabits per Second |
| MCDR | Mission Critical Design Review |
| MCR | Mission Concept Review |
| MCT | Mercury Cadmium Telluride |
| MDR | Mission Definition Review |
| MEL | Master Equipment List |
| MMS | Magnetospheric Multi-Scale Mission |
| MOC | Mission Operations Center |
| MOI | Moment of Inertia |
| MOU | Memorandum of Understanding |
| MPDR | Mission Preliminary Design Review |
| Mpix | Mega pixels |
| NAS | National Academy of Sciences |
| NASA | National Aeronautics and Space Administration |
| NASTRAN | NASA Stress Analysis |
| NEAR | Near Earth Asteroid Rendezvous |
| NHA | Next Higher Assembly |
| NiCd | Nickel Cadmium |
| NiH | Nickel Hydride |
| NIR | Near Infrared |
| NIRCam | Near Infrared Camera |
| NOAO | National Optical Astronomical Observatory |
| NRC | National Research Council |
| NVR | Non-volatile Residue |
| OSIM | Operating System Interface Module |
| OTA | Optical Telescope Assembly |
| P/Y | Pitch and Yaw |
| Pan-STARRS | Panoramic Survey Telescope and Rapid Response System |
| PDR | Preliminary Design Review |
| Photo-z | Photometric Redshift-1 |
| PLA | Payload Adapter |
| PM | Primary Mirror |
| PM/SE/MA | Project Management/Systems Engineering/Mission Assurance |
| PRT | Platinum Resistance Thermometer |
| PSF | Point Spread Function |
| PSP | Participating Scientist Program |
| PSR | Pre-Ship Review |
| PWB | Printed Wiring Board |
| QE | Quantum Efficiency |
| RFI | Request for Information |
| RMS | Root Mean Square |
| RT | Real-time |
| RTD | Resistance Thermal Device |
| Rx | Prescription |
| RY | Real Year |
| S/C | Spacecraft |
| S/N | Signal/Noise |
| SAA | South Atlantic Anomaly |
| SCA | Sensor Chip Assembly |
| SDO | Solar Dynamics Observatory |
| SDRAM | Synchronous Dynamic Random Access Memory |
| SDSS | Sloan Digital Sky Survey |
| SF | Super Field |
| Si | Silicon |
| SIR | Systems Integration Review |
| SLOC | Single Lines of Code |
| SM | Secondary Mirror |
| SN | Supernova |
| SNAP | Supernova Acceleration Probe |
| SNe | Supernovae |
| SOC | Science Operations Center |
| SpC | Spectrometer channel |
| SPOT | Satellite Pour l'Observation de la Terre |
| SPSO | Science Proposal Support Office |
| SpW | Space Wire |
| SRR | System Requirements Review |
| SSC | Science Support Center |
| SSR | Solid State Recorder |
| STEREO | Solar Terrestrial Relations Observatory |
| STScI | Space Telescope Science Institute |
| SZ | Sunyaev-Zeldovich |
| TBD | To be determined |
| Tbits | Terabits |
| TIRS | Thermal Infrared Scanner |
| TIS | Teledyne Imaging Sensors |
| TJGaAs | Triple Junction Gallium Arsenide |
| TM | Tertiary Mirror |
| TMA | Three Mirror Anastigmat |
| TRL | Technology Readiness Level |
| TRMM | Tropical Rainfall Measuring Mission |
| TWTA | Traveling Wave Tube Amplifier |
| UCB | University of California - Berkley |
| UH | University of Hawaii |
| US | United States |
| vDC | Volts Direct Current |
| VIS | Visible |
| WFC | Wide Field Camera |
| WFC3 | Wide Field Camera 3 |
| WL | Weak Lensing |
| WMAP | Wilkinson Microwave Anisotropy Probe |
| XTE | X-ray Timing Explorer |
| ZnSe | Zinc Selenide |